\documentclass[sigconf]{acmart}
\usepackage{lipsum}
\usepackage{array}
\usepackage{subcaption} 
\usepackage{multirow}
\usepackage{makecell}
\usepackage{fontawesome}
\usepackage{tabularx}
\usepackage{graphicx}
\usepackage{placeins}
\usepackage{booktabs}
\usepackage[export]{adjustbox}
\usepackage{cellspace}
\usepackage{adjustbox}
\usepackage{tabularray}
\usepackage{footnote}
\usepackage[capitalize]{cleveref}
\usepackage{multirow}
\usepackage[multiple]{footmisc}
\usepackage{hyperref}
\usepackage{enumitem}
\usepackage[normalem]{ulem}
\useunder{\uline}{\ul}{}

\AtBeginDocument{%
  }

\setcopyright{acmlicensed}
\copyrightyear{2025}
\acmYear{2025}
\acmDOI{XXXXXXX.XXXXXXX}

\definecolor{vividauburn}{rgb}{0.58, 0.15, 0.14}
\definecolor{rufous}{rgb}{0.66, 0.11, 0.03}
\definecolor{red-brown}{rgb}{0.65, 0.16, 0.16}
\definecolor{red(ncs)}{rgb}{0.77, 0.01, 0.2}

\author{Tianrun Qiu}
\orcid{0009-0002-9878-663X}
\authornote{Equal contribution.}
\email{qiutr@mail.sustech.edu.cn}
\affiliation{%
  \department{Department of Computer Science and Engineering}
  \institution{Southern University of Science and Technology}
  \city{Shenzhen}
  \state{Guangdong}
  \country{China}
}

\author{Changxin Chen}
\authornotemark[1]
\email{chenchangxin2022@mail.sustech.edu.cn}
\affiliation{%
  \department{Department of Computer Science and Engineering}
  \institution{Southern University of Science and Technology}
  \city{Shenzhen}
  \state{Guangdong}
  \country{China}
}

\author{Sizhe Cheng}
\email{sizhe003@e.ntu.edu.sg}
\affiliation{%
  \department{College of Computing and Data Science}
  \institution{Nanyang Technological University}
  \city{Singapore}
  \country{Singapore}
}

\author{Xuyang Liu}
\email{12310606@mail.sustech.edu.cn}
\affiliation{%
  \department{Department of Computer Science and Engineering}
  \institution{Southern University of Science and Technology}
  \city{Shenzhen}
  \state{Guangdong}
  \country{China}
}

\author{Xumeng Wang}
\email{wangxumeng@nankai.edu.cn}
\affiliation{%
  \department{DISSec}
  \institution{Nankai University}
  \city{Tianjin}
  \country{China}
}

\author{Zhicong Lu}
\orcid{0000-0002-7761-6351}
\email{zlu6@gmu.edu}
\affiliation{
  \department{Department of Computer Science}
  \institution{George Mason University}
  \city{Fairfax}
  \state{Virginia}
  \country{USA}
}

\author{Yuxin Ma}
\authornote{Corresponding author.}
\orcid{0000-0003-0484-668X}
\email{mayx@sustech.edu.cn}
\affiliation{
  \department{Department of Computer Science and Engineering}
  \institution{Southern University of Science and Technology}
  \city{Shenzhen}
  \state{Guangdong}
  \country{China}
}

\begin{document}

\title[GamerAstra]{GamerAstra: Supporting 2D Non-Twitch Video Games for Blind and Low-Vision Players through a Multi-Agent Framework}



\begin{abstract}
Blind and low-vision (BLV) players face critical challenges in engaging with video games due to the inaccessibility of visual elements, difficulties navigating interfaces, and limitations in performing interaction. Meanwhile, the development of specialized accessibility features typically requires substantial programming effort and is often implemented on a game-by-game basis. To address these challenges, we introduce GamerAstra, a multi-agent human-AI collaboration framework that leverages a multi-agent design to facilitate access to 2D non-twitch video games for BLV players. It integrates vision-language models and computer vision techniques, enabling interaction with games lacking native accessibility support. The framework also incorporates custom assistance granularities to support varying degrees of visual impairment and enhances interface navigation through multiple input modalities. Technical evaluations and user studies indicate that GamerAstra effectively enhances playability and provides a more immersive gaming experience for BLV players. These findings also underscore potential avenues for advancing intelligent accessibility frameworks in the gaming domain.
\end{abstract}

\ccsdesc[500]{Human-centered computing~Accessibility systems and tools}

\keywords{Accessibility, Blind, Visually Impaired, Low-Vision, Assistive Technology, Gaming, Vision-Language Model, Multi-Agent Framework}
\begin{teaserfigure}
  \centering
  \includegraphics[width=0.92\textwidth]{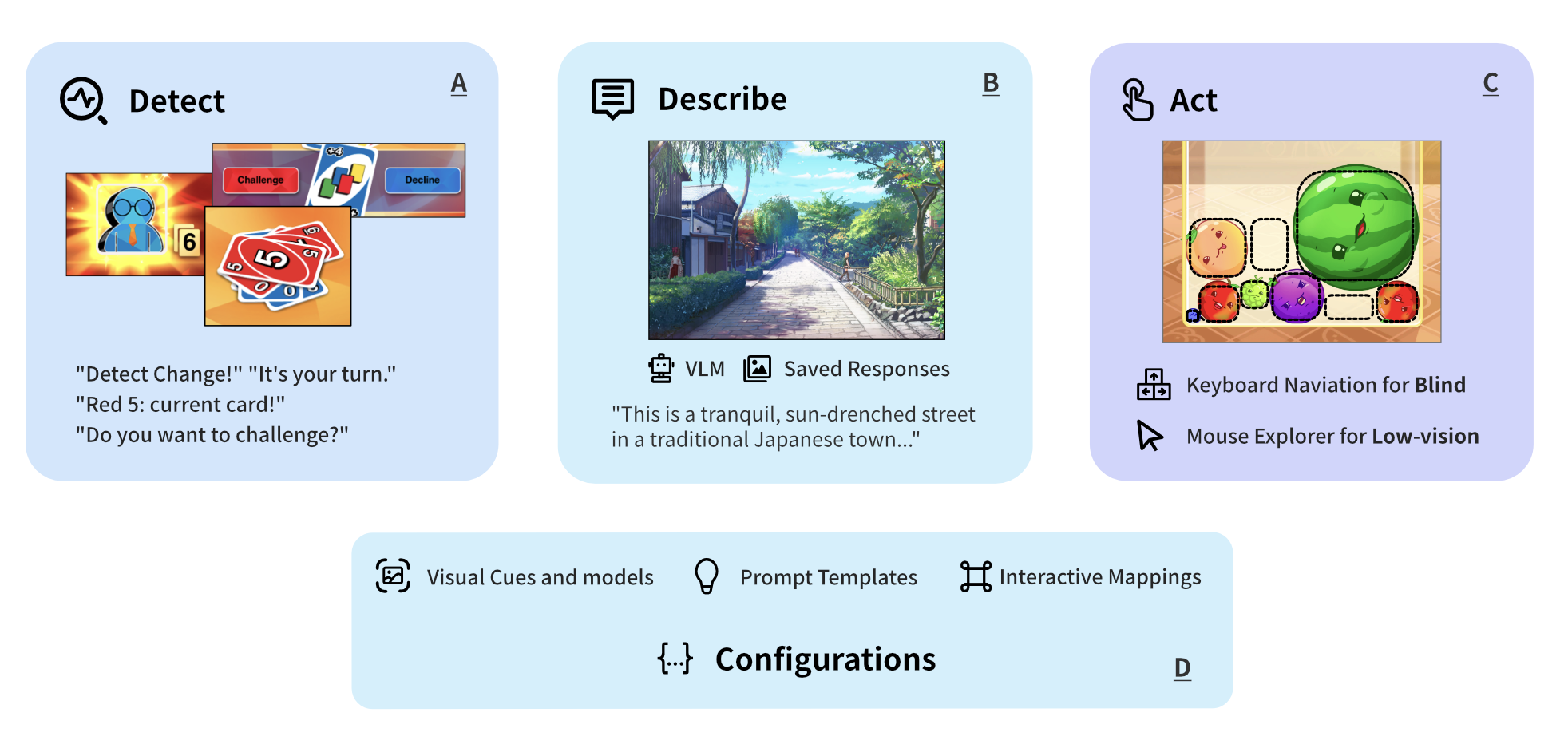}
  \caption{GamerAstra is a generalizable multi-agent framework that makes 2D non-twitch video games accessible for blind and low-vision (BLV) players, enabling new games to be integrated without requiring programming or technical expertise. (A) The Detect agent identifies critical in-game changes and interface states, providing timely cues for game-play awareness and potentially triggering the operations of other agents. (B) The Describe agent leverages vision-language models and pre-saved responses to produce rich contextual descriptions of environments and objects. (C) The Act agent facilitates interaction by offering accessible input methods tailored to different degrees of visual impairment, including keyboard navigation for blind players and mouse exploration for low-vision players. (D) A configuration system supports the coordinated operation of these three agents, including visual cues, prompt templates, and interactive mappings, which allow the framework to flexibly adapt to diverse game genres and accessibility requirements.}
  \Description{This image presents a layout of four main sections, each in rounded rectangles, focused on tools for interacting with visual content, likely for accessibility or visual processing. The top row has three sections: Section A (labeled "Detect" with a waveform icon) includes game-related images (a character with a number 6, UNO cards, "Challenge/Decline" buttons) and text prompts like "Detect Change!" and "It is your turn." Section B (labeled "Describe" with a speech bubble icon) shows a serene sunlit street in a traditional Japanese town, with options "VLM" (robot icon) and "Saved Responses" (picture icon), plus the description "This is a tranquil, sun-drenched street in a traditional Japanese town...". Section C (labeled "Act" with a hand cursor icon) features colorful cute characters (like a large green watermelon) in a grid, with options "Keyboard Navigation for Blind" (keyboard icon) and "Mouse Explorer for Low-vision" (mouse icon). Below these, Section D (labeled "Configurations" with curly braces icon) includes three items above: "Visual Cues and models" (square icon), "Prompt Templates" (lightbulb icon), and "Interactive Mappings" (connector icon), indicating setup options for the other features.}
  \label{fig:teaser}
\end{teaserfigure}


\maketitle

\section{Introduction}

Video gaming has become a significant medium of entertainment in today's world~\cite{kaltman2017stabilization}. However, BLV individuals face substantial challenges in engaging with video games due to the lack of adequate accessibility features in many titles~\cite{andrade2019playing, gonccalves2020playing, archambault2007computer, porter}. Despite the publication of an accessibility guide for game development as early as 2004~\cite{igda2004accessibility}, a significant proportion of video games still do not support screen readers or other common assistance techniques, making them inaccessible for most BLV players~\cite{andrade2019playing, yuan2011game, martinez2024playing}. This lack of accessibility often leads to frustration among BLV players due to unsatisfactory gaming experiences and limited game choices~\cite{gonccalves2023my}. Additionally, not being able to enjoy diverse types of video games alongside their sighted peers can contribute to feelings of loneliness for some BLV players~\cite{gonccalves2020playing, EasiertoPlayAlone}.

Efforts to improve accessibility have been made from various perspectives, but the vast majority of video games still remain inaccessible to BLV players. On the positive side, some games now include built-in accessibility features for players with visual impairments, such as \textit{Uncharted 4}~\cite{10.1145/3569219.3569335} and \textit{The Last of Us Part II}~\footnote{\url{https://www.playstation.com/en-us/games/the-last-of-us-part-ii}}. Additionally, games that expose official APIs, such as \textit{Stardew Valley}~\footnote{\url{https://www.nexusmods.com/stardewvalley/mods/16205}} and \textit{Minecraft}~\footnote{\url{https://www.curseforge.com/minecraft/mc-mods/blind-accessibility}}, allow users or third-party developers to create game modifications (often called ``mods'') that enhance accessibility. However, the adaptation of games for BLV players remains slow due to the high technical complexity involved and the relatively small size of the BLV gaming community~\cite{yuan2011game}. Furthermore, while extensive research has been undertaken to optimize the gaming experience for BLV players~\cite{nair2024surveyor, smith2018rad, 10.1145/3597638.3608386}, the practical implementation of these advancements remains limited. The adoption of accessibility-related insights often hinges on the willingness and priorities of game developers and producers, who may face constraints such as insufficient attention to accessibility, limited budgets, or competing development priorities.




With the emergence of next-generation artificial intelligence (AI) models, particularly visual language models (VLMs), there is promise in exploring human-AI collaborative gaming assistance in the absence of official accessibility functionality support. However, although advanced AI techniques have been integrated into multiple domains~\cite{jain2023towards, chang2024worldscribe, sharif2022voxlens, hirabayashi2023visphoto, huh2024designchecker, zhang2022a11yboard, huh2023genassist} by the human-computer interaction (HCI) community to empower the BLV gamers, relatively little effort has been devoted to identifying their potential as copilots for assisting BLV individuals in enjoying video games.

Therefore, we introduce GamerAstra, a multi-agent human-AI collaboration framework designed to make 2D non-twitch video games accessible without requiring programming skills or first-party support from game developers, which integrates three specialized agents: the \textit{Detect} Agent, the \textit{Describe} Agent, and the \textit{Act} Agent. By dynamically monitoring game states, providing contextually rich auditory descriptions, and enabling non-visual interactions, GamerAstra empowers BLV players to engage autonomously with game interfaces to eliminate traditional accessibility barriers, as shown in Fig.~\ref{fig:teaser}.



To demonstrate the effectiveness of the framework, the evaluation comprises two components: (1) a within-subject study involving eight BLV participants, which compares performance from baseline support tools to that achieved with GamerAstra, and (2) five quantitative evaluations to test the framework's accuracy and adaptability. The results indicate that GamerAstra significantly enhances game playability, provides a more immersive gaming experience, and demonstrates robustness across multiple scenarios. Based on these findings, we explore future directions for developing an even more unified, intelligent, and fully real-time framework, which could further promote equity between BLV individuals and the general public within the gaming context. GamerAstra represents a step towards enabling BLV players to experience gaming on par with their sighted peers.

In summary, our contributions include the following:

\begin{itemize}[leftmargin=*]
    \item A multi-agent accessibility framework, \textit{GamerAstra}, that facilitates an inclusive gaming experience for BLV players without necessitating official game support or additional program development to enhance accessibility in game-play for BLV players.
    \item A user-centric evaluation which includes a comparative study with BLV participants to highlight significant gains in playability and overall gaming experience, as well as a quantitative assessment to demonstrate the performance and adaptability across different games.
\end{itemize}

\section{Related work}

Our work builds upon recent research on BLV gaming and AI agents. In this section, we review key studies relevant to our work.

\subsection{Accessibility Applications using Large Language Models}
With the rapid advancement of Large Language Models (LLMs) and Vision-Language Models (VLMs), along with their improved understanding capabilities, increasing research focuses on applying these technologies to enhance accessibility for BLV individuals. 

\subsubsection{Accessibility for visual content understanding}

Recent studies focus on leveraging these models to assist BLV individuals by providing detailed image descriptions, interpreting complex visual scenes, and enhancing access to visual information across various digital platforms. WorldScribe~\cite{chang2024worldscribe} offers context-aware live visual descriptions that progressively detail objects, while Immersive A/V~\cite{jain2023towards} makes sports broadcasts accessible through sonification. VoxLens~\cite{sharif2022voxlens} presents an interactive plug-in that leverages AI to make data visualizations more accessible. For videos, both NarrationBot-and-InfoBot~\cite{10.1145/3597638.3608402} and ShortScribe~\cite{Van_Daele_2024} focus on enhancing video content accessibility for BLV users, while DanmuA11y~\cite{Xu2025DanmuA11y} focuses on transforming visually-centric barrages into accessible audio discussions, although both of them require pre-processing. While the former introduces a tandem visual dialogue system to provide richer, more interactive video descriptions, the latter uses models to generate hierarchical descriptions of short-form videos, helping users better understand and select content. Comics, which also pose unique accessibility challenges, are addressed by DCC~\cite{kim_suhyun_2024} through dense captioning techniques. 

Collectively, these works showcase the expanding role of advanced AI models in transforming visual content into accessible formats, empowering BLV users to engage more fully with visual media. However, the characteristics of being real-time, interface complexity, and the unfamiliarity of VLM with virtual environments hinder the existing work's effectiveness in understanding games.

\subsubsection{Accessibility for performing visual tasks}

Beyond understanding visual content, LLMs have also been instrumental in enabling BLV users to perform tasks traditionally requiring sight. GenAssist~\cite{huh2023genassist} allows BLV individuals to generate images using diffusion models, illustrating how AI can act as a ``copilot'' in completing visually demanding tasks. For editing videos, LAVE~\cite{wang2024lave} leverages LLMs to assist BLV users with video retrieval, brainstorming, and sequencing. DesignChecker~\cite{huh2024designchecker} aids BLV web developers in creating visually appealing websites, while A11yBoard~\cite{zhang2022a11yboard} makes digital art-boards accessible through gesture operations and multi-modal feedback. Similarly, VizAbility~\cite{Gorniak2024VizAbility} enhances chart accessibility by integrating an LLM-based question-and-answer module, supporting BLV users to navigate and explore data more easily, while VisPhoto~\cite{hirabayashi2023visphoto} improves photography with omnidirectional imaging and AI post-processing. Additionally, VisionTasker~\cite{Song_2024} utilizes VLMs for mobile automation, though limited by slow speed and inefficacy in gaming operations.

These innovations show how AI tools increase BLV users' independence, aiding in creative, professional, and daily activities. However, current methods lack real-time interactivity, immersion, and complex operation support for gaming, compounded by the absence of a control tree and insufficient precise coordinate control in VLMs.

\subsection{AI Agents in Gaming}

The integration of AI agents in gaming has been explored through various approaches, particularly with the rise of large language models (LLMs). The concept of LLM-based Game Agents (LLMGA) is systematically outlined in a recent survey~\cite{hu2024surveylargelanguagemodelbased}, which categorizes different agent types and highlights their practical applications. Among these implementations, Voyager~\cite{wang2023voyager} demonstrates an agent capable of autonomously playing Minecraft using GPT-4~\cite{gpt-4}, although it remains constrained by its reliance on Minecraft's automation API. Besides that, Atari-GPT~\cite{waytowich2024atarigptbenchmarkingmultimodallarge} investigates the potential of multi-modal LLMs to function as low-level policy agents in Atari games, showcasing their ability to interpret game environments and make decisions.

Beyond these task-specific models, prompt-based gaming agents have been emerging. Cradle~\cite{tan2024cradle} exemplifies this approach by using visual input and translating them into keyboard and mouse operations, supporting various games such as \textit{Red Dead Redemption II}~\footnote{\url{https://www.rockstargames.com/reddeadredemption2}} and \textit{Stardew Valley}~\footnote{\url{https://www.stardewvalley.net/}}. However, its slow reaction time limits its usability outside of controlled research settings, and its complex code-based configuration makes external contributions rather difficult. De Wynter's work~\cite{de2024will} on playing \textit{DOOM} using prompts further reveals challenges related to spatial understanding and the complexity of prompt formulation. For complex action role-playing games (ARPGs), VARP~\cite{chen2024vlmsplayactionroleplaying} integrates vision inputs and task planning for complex actions in the AAA games \textit{Black Myth: Wukong}. However, its reliance on pre-recorded data and human guidance, along with high training demands, highlights ongoing challenges in developing adaptable and efficient game agents. In the meantime, general-usage GUI agents, e.g. CogAgent~\cite{hong2024cogagentvisuallanguagemodel} and Claude Computer-Use~\cite{claude-computer-use}, are mainly trained based on computer programs' UI, and have difficulty handling game scenarios and cannot satisfy real-time interaction requirement due to their slow speed.

\subsection{Games Designed or Optimized for BLV Population} \label{games-for-blv}

Efforts to make games accessible to BLV users have been diverse. Audio games, such as those listed on audiogames.net~\cite{audiogames}, cater specifically to BLV players. However, research shows that many BLV game players feel unable to integrate because they cannot play games with their sighted friends~\cite{gonccalves2020playing}. Fortunately, some video games~\footnote{\url{https://www.playstation.com/en-us/games/the-last-of-us-part-ii}}\footnote{\url{https://diablo4.blizzard.com/en-us}} have incorporated official accessibility features. Additionally, non-official mods for games like \textit{Stardew Valley}~\footnote{\url{https://www.nexusmods.com/stardewvalley/mods/16205}}, \textit{Minecraft}~\footnote{\url{https://www.curseforge.com/minecraft/mc-mods/blind-accessibility}} and \textit{Hearthstone}~\footnote{\url{https://hearthstoneaccess.com}} further enhance accessibility. These adaptations demonstrate a growing recognition of the need for inclusive gaming experiences, although challenges remain in creating games that appeal equally to sighted and BLV players. However, due to cost and profit limitations, there is still a lack of high-quality productions that sufficiently meet the BLV community's recreational needs, compared to the video game market's prosperity.

\section{Challenges and Motivation}

In this section, we first enumerate the key challenges distilled from the literature discussed in the previous section, then describe systemic limitations that motivate our approach, and finally outline how these motivations drive the design of our proposed framework.

\subsection{Challenges}

Existing research has contributed significantly to understanding BLV gaming experiences. The RAD~\cite{smith2018rad} and NavStick~\cite{nair2021navstick} provide audio-based tools for navigating virtual environments and racing games, while Surveyor~\cite{nair2024surveyor} facilitates exploration in games. We synthesise these barriers into six concrete challenges:

\begin{enumerate}
    \item \textbf{Inaccessible game UI components.} These include screen reader incompatibility~\cite{ran2025usersblindlowvision, andrade2020introducing} and inaccessible menus~\cite{gonccalves2023my, martinez2024playing}, which can cause psychological distress~\cite{ran2025usersblindlowvision} and reduce player independence~\cite{gonccalves2023my}.
    
    \item \textbf{Lack of real-time feedback.} Players must memorize game states and environments, use compensatory strategies like resource estimation through incremental actions, and cope with missing timely auditory cues for critical events~\cite{gonccalves2023my, gonccalves2020playing}.
    
    \item \textbf{Difficulties in precise operations.} Tasks such as aligning with doors or NPCs often require inefficient trial-and-error or memorization, leading to frustration for BLV players~\cite{gonccalves2023my}.
    
    \item \textbf{Barriers in accessing complex visual elements.} Uninterpretable cues obscure objectives and progress, hindering onboarding and forcing reliance on external guides~\cite{gonccalves2023my}; complex graphics in mainstream games worsen exclusion~\cite{ran2025usersblindlowvision}, while oversimplified BLV-specific alternatives often lack meaningful depth, reducing engagement~\cite{gonccalves2020playing}.
    
    \item \textbf{Excessive assistance and verbose feedback.} Features like auto-play remove strategic elements and may feel patronizing~\cite{ran2025usersblindlowvision}, and fixed assistance systems ignore diverse player preferences, highlighting the need for customizable opt-in features~\cite{martinez2024playing}.
    
    \item \textbf{Insufficient support for low-vision players.} Many BLV-specific games and accessibility plugins (e.g., Minecraft, Hearthstone) prioritize auditory solutions for legally blind players, leaving low-vision players without tailored support across different impairment levels~\cite{othman2023serious, ran2025usersblindlowvision}.
    
\end{enumerate}

However, to address these challenges and ensure that the relevant efforts are implemented effectively, the workload of game developers remains indispensable to apply these existing ideas, while in reality, the adaptation of games for BLV players has always been slow, given low profitability and high cost~\cite{yuan2011game}. Other work on accessible gaming includes Accessible Play~\cite{10.1145/3573382.3616075}, which uses the interactive process modeling method to analyze the current situation of accessibility supports. Meanwhile, a developer-oriented framework~\cite{10.1007/978-3-031-76818-7_7} has been proposed to evaluate the accessibility of existing games. However, these efforts are still primarily focused on the developer side and remain in the early stages of implementation, with few efforts addressing retroactive accessibility. As such, practical applications that directly benefit BLV players are yet to be fully implemented. 

\subsection{Motivation: Bridging the Gap}

Although existing work identifies BLV gaming challenges, three systemic limitations hinder progress:

\begin{itemize}
    \item \textbf{Genre constraints:} BLV players are often confined to simple audio-based or turn-based games~\cite{gonccalves2023my}, lacking the complexity and appeal of games played by sighted people~\cite{gonccalves2020playing} while excluding them from the mainstream gaming culture and peer interactions~\cite{gonccalves2020playing}.

    \item \textbf{Technical and economic barriers:} While some mods (e.g., \textit{Stardew Valley Accessibility}) offer partial solutions, they require programming expertise and community-specific knowledge~\cite{gonccalves2023my}, limiting their reach. Game studios, especially those in the Global South, often hesitate to implement native accessibility features due to development costs and relatively niche markets, while game engine support remains insufficient~\cite{yuan2011game, ran2025usersblindlowvision}. As a result, most games, especially legacy games, lack accessibility features and cannot be read by screen reader software, despite the strong desire of BLV players for mainstream integration~\cite{martinez2024playing}.

    \item \textbf{Design rigidity:} Current solutions lack adaptability, as verbose feedback disrupts immersion~\cite{andrade2020introducing}, while ``one-size-fits-all'' approaches cannot meet the needs of low-vision players~\cite{ran2025usersblindlowvision}.
\end{itemize}
 
The emergence of advanced AI, particularly vision-language models (VLMs), offers a promising paradigm: VLMs can act as ``copilots'' that interpret visuals, reduce cognitive load, and enable richer interactions in games~\cite{ran2025usersblindlowvision, agrimi2024game}. Nonetheless, existing work has not yet produced a system that both realizes this potential and simultaneously addresses the systemic limitations identified above. 


\subsection{From Motivation to Framework}
Guided by the challenges and insights outlined above, we propose a \emph{no-code, multi-agent, augmentable} framework that \textbf{(1)} can be applied retroactively to a wide range of titles, \textbf{(2)} uses VLM-based copilots to interpret visual scenes and generate concise, context-aware feedback, and \textbf{(3)} exposes configurable policies and plugins so assistance is opt-in and fine-grained. We initially target 2D non-twitch games where current model latency is compatible with interactive play; however, the framework is explicitly designed to be modular so that more demanding genres and native integrations can be supported as tooling and industry incentives evolve. By implementing the goals above, our approach aims to reduce developer burden, increase BLV participation in gaming, and preserve the depth and agency that make games meaningful.

\section{GamerAstra}

\begin{figure}
    \centering
    \includegraphics[width=1\linewidth]{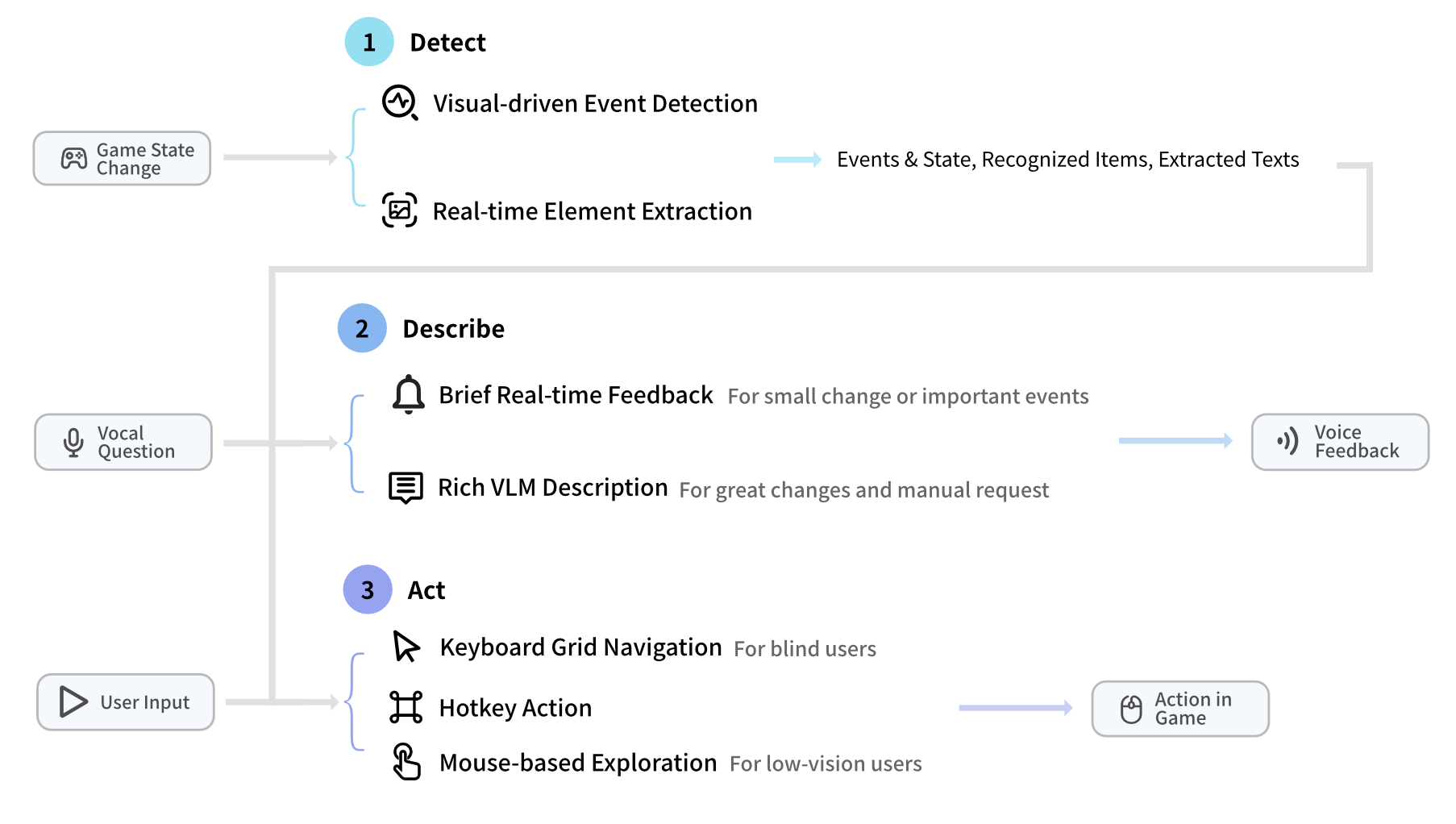}
    \caption{The three agents within the framework and their interactions}
    \label{fig:agents-reaction}
    \Description{This image is a diagram, outlining a system with three main stages: Detect, Describe, and Act. The Detect stage (marked with a blue circle labeled "1") has two components: "Visual-driven Event Detection" (with a magnifying glass icon) and "Real-time Material Extraction" (with a square icon), taking "Game State Change" as input and producing "Events \& State, Recognized Items, Extracted Texts" as output. The Describe stage (blue circle "2") includes "Brief Real-time Feedback" (bell icon, for small/important events) and "Rich VLM Description" (speech bubble icon, for major changes/manual requests), with "Vocal Question" as input and "Voice Feedback" as output. The Act stage (blue circle "3") features "Keyboard Grid Navigation" (cursor icon, for blind users), "Hotkey Action" (keyboard icon), and "Mouse-based Exploration" (mouse icon, for low-vision users), using "User Input" as input to generate "Action in Game" as output. The diagram illustrates how these stages interact to enable a system, likely assisting visually impaired users in engaging with games via detection, description, and interaction.}
\end{figure}

To provide better assistance to video games without adequate accessibility support, we propose \textbf{GamerAstra}, a human-AI collaboration framework for assisting game-play for BLV players, which comprises three distinct agents: a \textbf{Detect} Agent, a \textbf{Describe} Agent, and an \textbf{Act} Agent, each endowed with multiple functional capabilities. This allocation of workload enhances the efficiency of information retrieval and processing, allowing for real-time assistance tailored to the needs of BLV players, with their relationship shown in Fig.~\ref{fig:agents-reaction}. 

\subsection{Detection}

The \textit{GamerAstra} framework employs the \textbf{Detect} agent as its initial point of interaction. This agent continuously monitors the game state for critical changes and dynamically extracts essential information. It functions with low latency to provide immediate utility to BLV players while automatically triggering and configuring subsequent agents in framework as required. Its core functionalities include:

\begin{itemize}[leftmargin=*]
    \item \textbf{Visually-Driven Event Detection} identifies significant visual cues, important items and game state transitions presented on the interface constantly using real-time models and algorithms. Upon detection, it is provided to subsequent agents to notify BLV players with real-time auditory announcements of the relevant events. This functionality also automatically signals the framework to update contextually appropriate navigation guidance and hotkey lists, enabling BLV players to build an accurate mental map of the game environment and facilitating self-directed exploration.
    \item \textbf{Realtime Element Extraction} continuously polls the game-play interface to dynamically capture essential textual content and identify specific in-game items. This extracted information, kept current with the game state, is furnished to subsequent stages of the framework. This ensures BLV players receive the critical, real-time elements necessary for informed interaction and decision-making during game-play.
\end{itemize}

Crucially, configuring the \textbf{Detect} agent's capabilities requires no programming expertise. In addition, community volunteers, acting as external contributors, can provide straightforward resources (e.g., interface reference screenshots, item exemplars) to facilitate event detection and element extraction. Detailed configuration specifications are provided in Section~\ref{game-config-struct}.

\begin{figure*}
    \centering
    \includegraphics[width=1\textwidth]{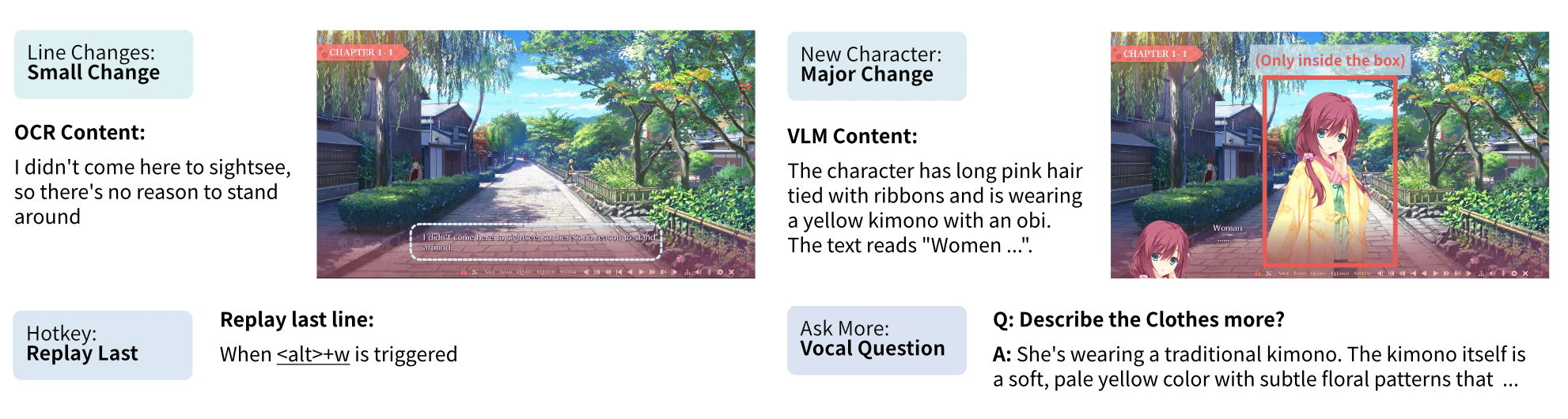}
    \caption{Multiple types of feedback in \textit{SenrenBanka} (a visual novel game), including responses of line changes, appearing of new characters, hotkeys and additional questions.}
    \label{fig:multi-feedback}
    \Description{This image is split into two main sections. On the left, under "Line Changes: Small Change," there’s OCR content reading "I didn't come here to sightsee, so there's no reason to stand around," paired with a game scene of a scenic path. A "Hotkey: Replay Last" section explains that pressing <alt>+w replays the last line. On the right, under "New Character: Major Change," VLM content describes a character with long pink hair in a yellow kimono, with the game scene now showing that character (framed in red). Below, "Ask More: Vocal Question" includes a question "Describe the Clothes more?" and an answer detailing the kimono's pale yellow, floral pattern. The image showcases how the game provides feedback for text changes, new characters, hotkey functionality, and vocal inquiries to assist users, likely those with visual impairments, in engaging with the game.}
\end{figure*}

\subsection{Description}

The next agent is the \textbf{Describe} Agent, which delivers contextually appropriate descriptions at precisely timed moments, empowering BLV players to explore visual elements, integrate fully into game narratives, and experience intended emotional responses. It provides dynamic scene narration alongside real-time interpretive feedback, enabling users to construct and maintain a comprehensive mental model of the game environment. When uncertainties arise during game-play, the agent can also function as an interactive conversational partner, resolving player queries through natural dialogue and thereby surpassing traditional unidirectional accessibility approaches.

To balance informativeness and clarity, the system automatically tailors its descriptive output using multi-tiered criteria, as described in Fig.~\ref{fig:multi-feedback}:

\begin{itemize}[leftmargin=*]
    \item \textbf{Ambient Silence for Non-Essential Changes}: The system intelligently suppresses auditory output during minor visual fluctuations that convey no meaningful game-play information, maintaining an unobtrusive experience. 
    \item \textbf{Concise Updates for Transient Information}: For rapidly changing game states requiring immediate attention, streamlined auditory feedback is delivered instantaneously with information gained by the Detect agent. In addition, the Describe agent will broadcast preset notifications when the Detect agent detects important events for real-time acknowledgment.
    \item \textbf{Vivid Narration for Significant Changes}: Major contextual developments trigger rich, focused descriptions prioritizing novel elements from VLM while explicitly avoiding redundant information from prior states. 
\end{itemize}

Such adaptive approach ensures that all auditory feedback remains contextually relevant and temporally optimized. Meanwhile, players can retain full control: critical notifications can interrupt ongoing narration, descriptions can be manually paused or resumed using hotkeys, and suppressed content remains available for later review.

BLV players maintain autonomous exploration capabilities through two distinct inquiry methods:

\begin{itemize}[leftmargin=*]
    \item \textbf{Natural Language Dialogue}: Players can verbally request specific information at any moment using vocal queries, receiving precise context-aware responses that emulate human co-pilot assistance. 
    \item \textbf{Contextual Hotkey Queries}: Programmable actions provide immediate access to frequently needed game state details (e.g., recent events or current status indicators), as detailed in Section~\ref{hotkey-actions} later. Depending on the hotkey configuration, instant feedback can be obtained or VLM can be invoked to get detailed feedback.
\end{itemize}

\subsection{Action}

Performing spatially-dependent actions (e.g., button selection, menu navigation) presents significant challenges for BLV players when operating visually complex game interfaces. Our framework overcomes these barriers by introducing the \textbf{Act} agent, which enables the execution of more precise interactions without reliance on visual targeting or memorized sequences, empowering BLV players to independently perform essential game-play actions. 

To utilize the information obtained from these sources, two methods are proposed for general user interaction, which are \textbf{keyboard arrow keys navigation} and \textbf{hotkey actions} to cater to the specific and divergent needs of different players on different occasions, as shown in Fig.~\ref{fig:navigation}. Low vision players can also use the mouse for \textbf{mouse-based interactive interface exploration}.

\subsubsection{Keyboard Grid Navigation} 
Provides real-time spatial navigation through a dynamically structured representation of all active interface elements. Players explore interactive elements using arrow keys in two dimensions, receiving auditory position updates (e.g., "Row 3 of 5, Column 2 of 8") that reinforce spatial understanding while maintaining concise feedback. The layout preserves spatial relationships in the original interface to strengthen mental mapping. The system automatically prioritizes relevant elements during navigation, eliminating non-interactive distractions.

\subsubsection{Mouse-based Exploration}
For low-vision players, the same grid remains, but for them, using the mouse to obtain the corresponding element's description will be more efficient.

\subsubsection{Programmable Hotkeys} \label{hotkey-actions}

Preconfigured key bindings enhance game-play efficiency while maintaining narrative immersion by compressing multi-step operations into efficient single-action commands. Hotkey assignments are contextually aware, activating only during applicable game-play scenarios to prevent accidental triggering based on the context provided by the Detect agent, and can serve distinct interaction purposes:

\begin{itemize}[leftmargin=*]
    \item \textbf{Context-Sensitive Description}: 
    This enables detailed auditory exposition of specific screen regions when players require deeper environmental understanding. Upon activation, the system generates rich descriptive narratives focused solely on designated areas, elucidating spatial relationships between objects, character expressions, or subtle environmental details.
    
    \item \textbf{Interface Control Execution}: 
    Essential interface interactions, like menu confirmations, dialog selections, or system commands, can be abstracted into single-command executions to allow BLV players to maintain game-play momentum during time-sensitive operations, and eliminate coordination barriers.
    
    \item \textbf{Game State Information Retrieval}: 
    This feature enables BLV players to obtain the latest updates in real time whenever needed, or retrieve previously missed information to support strategic decision-making and situational awareness. Vital statistics may include health metrics, resource counts, timer values, and progression markers and are presented through concise auditory announcements.
\end{itemize}

\begin{figure*}[htbp]
    \centering
    \includegraphics[width=0.8\textwidth]{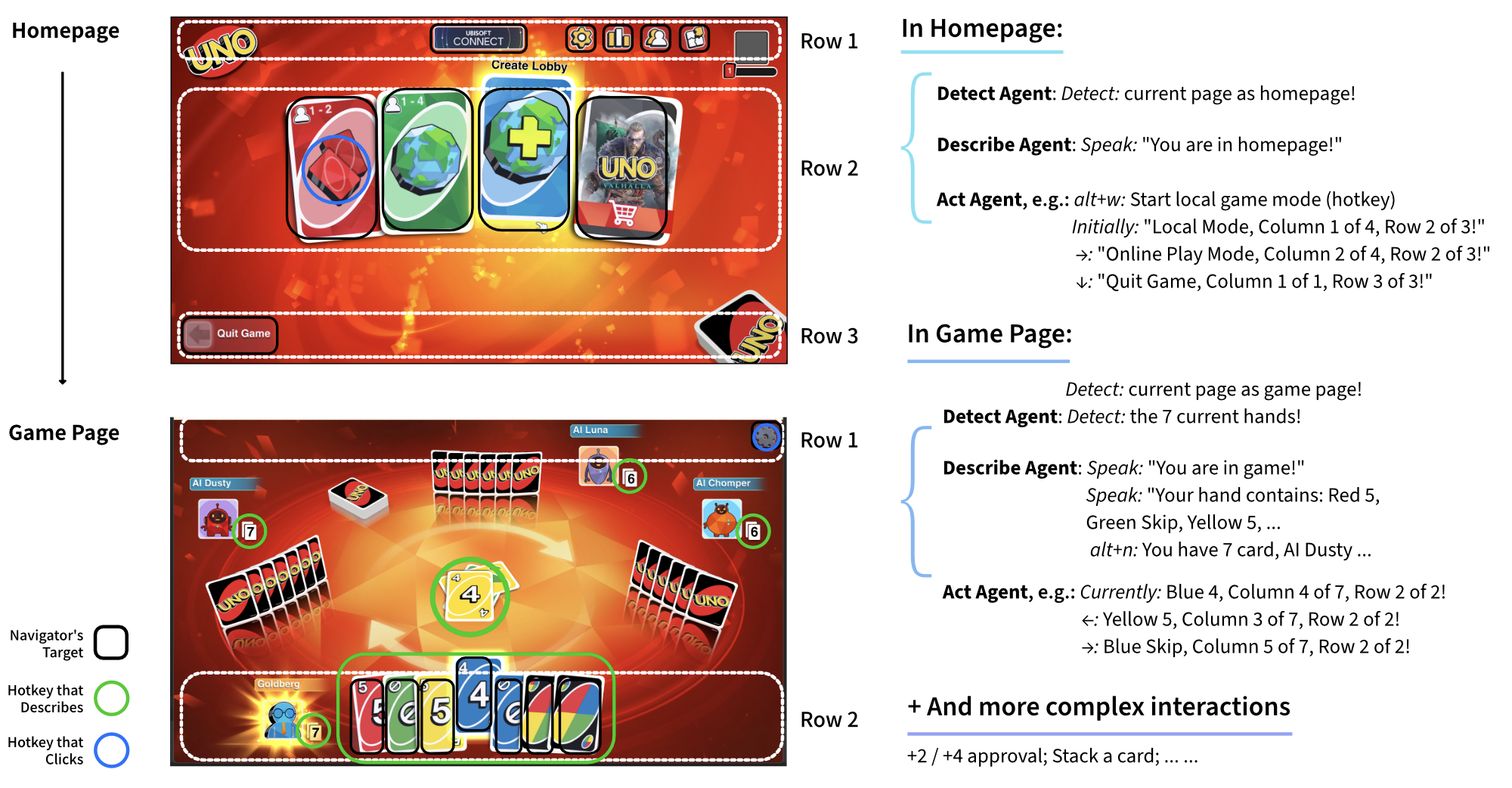}
    \caption{The use case in Uno game. \textit{Goldberg} is the default player name in Uno and is not related to authors.}
    \label{fig:navigation}
    \Description{This image illustrates a use case for the Uno game, featuring two main pages: the Homepage (index page with game mode options) and the Game Page (active game with cards and AI players). On the left, a vertical flow connects the Homepage to the Game Page, alongside a legend defining "Navigator's Target" (white square), "Hotkey that Describes" (green circle), and "Hotkey that Clicks" (blue circle). On the right, each page's functionality is explained via three agents: Detect, Describe, and Act. For the Homepage, the Detect Agent identifies the page as the homepage, the Describe Agent provides voice feedback ("You are in homepage!"), and the Act Agent uses hotkeys (e.g., alt+w) to switch game modes (with examples like "Local Mode, Column 1 of 4, Row 2 of 3!"). For the Game Page, the Detect Agent recognizes the page as a game page and identifies 7 current hands, the Describe Agent gives voice feedback ("You are in game!") and details the player’s hand (e.g., "Red 5, Green Skip, Yellow 5, ..."), and the Act Agent manages card interactions (with examples like "Currently: Blue 4, Column 4 of 7, Row 2 of 2!"). Additionally, there’s mention of more complex interactions such as +2/+4 approval and stacking cards.}
\end{figure*}

To summarize, these mechanisms collectively provide BLV players with efficient and self-directed control over game-play interactions, effectively eliminating the traditional trade-off between action speed and spatial understanding.

Due to technical limitations, although some interactive elements can be automatically detected, the specific interactive areas and whether a particular element is interactive often still require manual annotation, with manual refinement mechanisms available to contributors for guaranteeing absolute reliability. The detailed annotation workflow is described in Section~\ref{game-config-struct}.

\subsection{Usage Scenario}

To demonstrate the performance of our framework and the aforementioned Agents, we present a comprehensive game-play scenario featuring David, a blind player engaged in a digital UNO session. This scenario illustrates how GamerAstra's multi-agent system can enhance accessibility for BLV players, as shown in Fig.~\ref{fig:navigation}.

\begin{figure}
    \centering
    \includegraphics[width=0.98\linewidth]{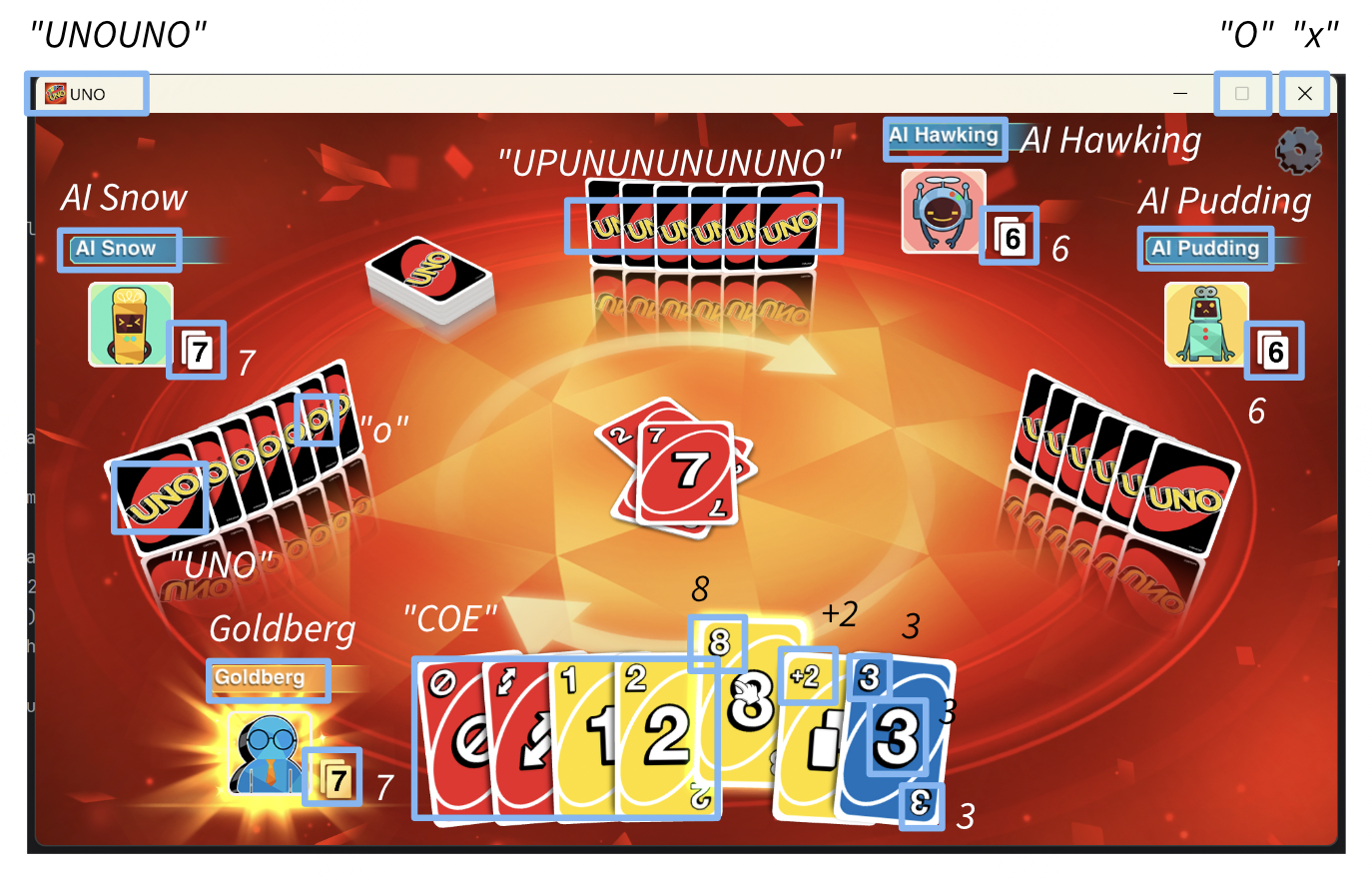}
    \caption{OCR result when using Paddle OCR.}
    \label{fig:ocr-result}
    \Description{This image shows a UNO game interface with non-optimal overlaid OCR results merely using Paddle OCR, in an inaccessible game. The game screen features multiple players: AI Snow (with 7 cards), AI Hawking (with 6 cards), AI Pudding (with 6 cards), and a player named Goldberg (with 7 cards). The central play area displays a red 7 card. Various text elements are detected and highlighted with blue boxes: at the top, texts like "UNOUNO" and "UPUNUNUNUNUNO"; near the window’s close/minimize buttons, "O" and "X"; by AI Snow’s cards, "UNO" and "O"; by Goldberg's cards, "UNO" and "COE"; and on the playable cards, numbers (1, 2, 8, +2, 3) and the UNO logo. }
\end{figure}

UNO is a classic card game that involves matching cards based on their color or number. Players take turns playing cards from their hand that match the top card in the central pile. The objective is to be the first to play all cards in hand. For BLV players, the visual nature of the game presents significant challenges, and the game is not playable when merely using OCR, as shown in Fig.~\ref{fig:ocr-result}.

\subsubsection{Homepage} 
After opening the Uno game, David first enters a homepage featuring multiple image-based buttons without persistent text cues, which is completely inaccessible to OCR, and thus incomprehensible to David. Therefore, the \textit{Detect Agent} is utilized to perform initial event detection, identifying the current page as the homepage through visual cues. The \textit{Describe Agent} immediately notifies David and instructs the \textit{Act Agent} to load the corresponding navigation grid for the homepage. Subsequently, David can select his desired mode using the keyboard to start the game, which is Local Mode in this case, while his low-vision friends can receive instant voice prompts when hovering the mouse over a button to them to choose a mode.

\subsubsection{Game Page}
Upon entering the game scene, the framework's \textit{Detect Agent} immediately identifies the transition from the system menu to the game-play interface through real-time visual state monitoring. It captures David's initial hand of seven cards using in-game item recognition templates, while simultaneously extracting player identifiers and turn indicators via OCR. This information is passed to \textit{Describe Agent}, which generates an auditory summary: ``You are Player Goldberg. Your hand contains: Red 5, Green Skip, Yellow 5, Blue 4, Blue Skip, Wild Card, Wild Card.''

When David's turn begins, the \textit{Detect Agent} recognizes the glowing halo flickering around the avatar frame through continuous state classification, triggering the \textit{Describe Agent} to announce ``Your turn'' and ``Last discard: Yellow 4.'' Then, he can navigate his hand using arrow keys to move focus between cards while receiving positional feedback (e.g., ``Column 4: Blue 4''). Upon selecting a card with $\text{SPACE}$, the \textit{Act Agent} executes the play action at the correct coordinate noted by the \textit{Detect Agent} previously. Despite that, if he did not hear the card at the top of the current pile clearly before, he can also use a shortcut to retrieve it again; and he can also use a shortcut to obtain information about which cards each user currently holds. Meanwhile, his friends with low vision can hover the mouse over various cards to obtain information about each one.

When encountering various other situations, GamerAstra can also assist David in resolving them. Throughout the game-play, the GamerAstra framework ensures the player receives timely, relevant feedback, and is capable of making moves, maintaining a seamless and immersive UNO experience.
\section{Implementation}

\begin{figure}
    \centering
    \includegraphics[width=1\linewidth]{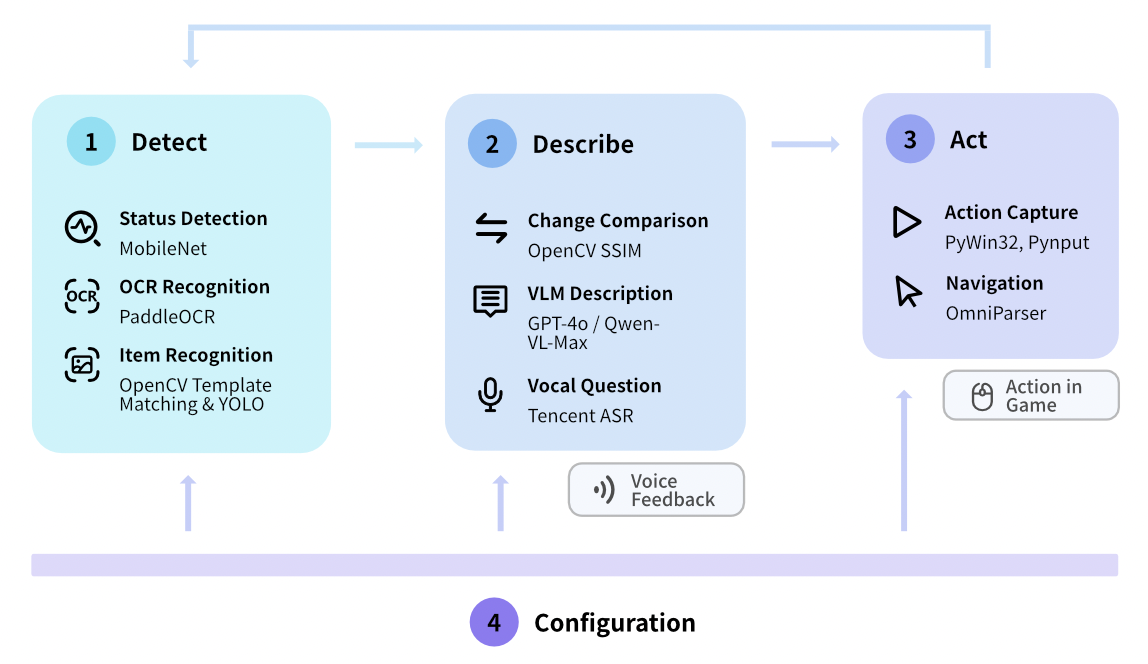}
    \caption{Implementation of agents}
    \label{fig:agents-implementation}
    \Description{This diagram outlines the implementation of three agents (Detect, Describe, Act) and configuration part in GamerAstra, for assisting with game interaction. The "Detect" stage (light blue, labeled 1) uses MobileNet for status detection, PaddleOCR for OCR recognition, and OpenCV Template Matching plus YOLO for item recognition. The "Describe" stage (blue, labeled 2) performs change comparison with OpenCV SSIM, visual-language model (VLM) description using GPT-4o or Qwen-VL-Max, and processes vocal questions via Tencent ASR, outputting voice feedback. The "Act" stage (purple, labeled 3) captures actions with PyWin32 and PyInput, handles navigation with OmniParser, and results in actions within a game. The "Configuration" stage (purple circle, labeled 4) sits at the bottom, supporting the Detect, Describe, and Act stages to ensure the pipeline functions cohesively, showing how each component uses specific technologies to detect, describe, and act on game elements.}
\end{figure}

The GamerAstra framework is structured around three high-level agents: \textbf{Detect Agent}, \textbf{Describe Agent}, and \textbf{Act Agent}, each coordinating a set of supporting functionalities, and supported by a \textbf{Configuration system} that enables accessible game interaction for blind and low-vision players, with their implementation showed in Fig.~\ref{fig:agents-implementation}. Additionally, we present a quantitative evaluation of the framework, focusing on the performance of these core agents. 

\subsection{Implementation Details}
GamerAstra is implemented in Python~\footnote{\url{https://www.python.org/}} and adopts a modular, multi-threaded architecture. We will describe the implementation details of the three core agents, highlighting their individual functionalities and methods used to realize them.


\subsubsection{Detect Agent}
The Detect Agent is responsible for identifying and localizing salient visual elements within the game environment that are essential for navigation and decision-making. It comprises three core functionalities: \textbf{status detection}, \textbf{OCR recognition}, and \textbf{item recognition}.

\textbf{Status Detection: } Based on MobileNetV2~\cite{sandler2019mobilenetv2invertedresidualslinear}, this module identifies the current game state by comparing cropped screenshots against a predefined set of visual cues. The classification process operates at approximately 2ms per inference on an Intel i7-11700 CPU, enabling high-frequency polling for rapid detection of state transitions.

\textbf{OCR Recognition: } Utilizing PaddleOCR~\cite{du2020ppocrpracticalultralightweight}, this component extracts text from the game interface in real time, returning both the recognized textual content and its corresponding screen coordinates.

\textbf{Item Recognition: } For static elements, OpenCV’s template matching~\cite{opencv_library} is employed using cropped reference images. In contrast, dynamic or distorted elements are detected using Ultralytics’ YOLO framework~\footnote{\url{https://ultralytics.com}}, specifically the YOLOv11 model~\cite{yolo11_ultralytics}. This kind of recognition also provides accurate coordinates and bonding boxes for elements.

\subsubsection{Describe Agent}
The Describe Agent translates visual scenes into concise and context-aware verbal summaries. It includes four key functionalities: \textbf{change comparison}, \textbf{visual-language model (VLM) description}, \textbf{voice recognition} and \textbf{voice feedback}, all designed to deliver adaptive real-time narration aligned with game-play dynamics.

\textbf{Change Comparison: } This module employs the Structural Similarity Index (SSIM)~\cite{ssim} from OpenCV~\cite{opencv_library} to compare successive frames. Based on the computed similarity $\Delta$, it triggers different feedback pathways according to two predefined thresholds ($\theta_1$, $\theta_2$):
\begin{itemize}[leftmargin=*]
\item If $\Delta < \theta_1$: the visual change is negligible and ignored.
\item If $\theta_1 \leq \Delta < \theta_2$: brief feedback is generated using OCR and item recognition results from the Detect Agent.
\item If $\Delta \geq \theta_2$: the VLM description module is activated to generate a detailed summary.
\end{itemize}

\textbf{VLM Description: }This functionality calls the OpenAI API~\footnote{\url{https://github.com/openai/openai-python}} to invoke models such as GPT-4o~\cite{gpt-4o} and Qwen-VL-Max~\cite{Qwen-VL} for image-based description generation. Before generating new descriptions, the system checks a local database to avoid duplication. Newly generated descriptions are stored for future reference. 

\textbf{Voice Recognition: }User speech input is transcribed using Tencent Automatic Speech Recognition (ASR) service~\cite{TencentASR}, enabling voice-based queries and commands. A hot word glossary of common gaming terms is provided to further align textual feedback with game content.

\textbf{Voice Feedback: }Voice feedback is given back to players for description. The system further employs spatial audio for changes related to items with detectable coordinates. Horizontal localization is based on the inter-aural intensity difference, achieved through gain allocation in the left and right channels, and enhanced for greater distinctiveness via a non-linear algorithm; vertical localization is achieved through pitch variations and temporal delays in the vertical direction.

\subsubsection{Act Agent}
The Act Agent operationalizes the interpreted game state and user intent by executing concrete interactions. It consists of two main components: \textbf{action capture} and \textbf{navigation}, both of which are focused on spatially-aware UI interaction.

\textbf{Action Capture: } This module uses PyWin32~\cite{pywin32} and Pynput~\footnote{\url{https://github.com/moses-palmer/pynput}} to monitor real-time keyboard and mouse inputs.

\textbf{Navigation: } This functionality utilizes OmniParser~\cite{lu2024omniparserpurevisionbased} to convert game UI into structured semantic representations, integrating results from Detect Agent and manually refined UI components. These elements are composed into a 2D spatial layout to support keyboard-based navigation and interaction.

\begin{table*}[!htb]
\centering \begin{tabular}{|l|l|l|l|l|l|}
\hline
Game                         & Metric             & NVDA + OCR & General & Auto-adaptive         & Full Framework \\ \hline
\multirow{2}{*}{Uno}         & Understand         & \textit{2/6}        & 5/6     & 5/6                   & \textbf{6/6}            \\
                             & Execute       & \textit{2/6}        & 3/6     & 3/6                   & \textbf{6/6}            \\ \hline
\multirow{2}{*}{SenrenBanka} & Understand         & \textit{2/5}        & 5/5     & 5/5                   & \textbf{5/5}            \\
                             & Execute       & \textit{2/5}        & 3/5     & 3/5                   & \textbf{5/5}            \\ \hline
\multirow{2}{*}{MySuika}     & Understand         & \textit{1/4}        & 3/4     & 3/4                   & \textbf{4/4}            \\
                             & Execute       & \textit{0/4}        & 2/4     & 2/4                   & \textbf{4/4}            \\ \hline
\end{tabular}
\caption{Comparison of Modes on Successful Scenes}
\Description{This table titled "Comparison of Modes on Successful Scenes" evaluates five modes, that are NVDA + OCR, General, Self-learning, and Full Framework, across three games (Uno, SenrenBanka, MySuika) using two metrics: "Understand" and "Execute." For Uno, under "Understand," NVDA + OCR scores 2/6, General and Self-learning score 5/6, and Full Framework scores 6/6; under "Execute," NVDA + OCR scores 2/6, General and Self-learning score 3/6, and Full Framework scores 6/6. For SenrenBanka, "Understand" shows NVDA + OCR at 2/5, General and Self-learning at 5/5, and Full Framework at 5/5; "Execute" has NVDA + OCR at 2/5, General and Self-learning at 3/5, and Full Framework at 5/5. For MySuika, "Understand" has NVDA + OCR at 1/4, General and Self-learning at 3/4, and Full Framework at 4/4; "Execute" shows NVDA + OCR at 0/4, General and Self-learning at 2/4, and Full Framework at 4/4. Overall, the Full Framework mode achieves the highest success in both understanding and executing tasks across all three games, outperforming the other modes.}
\label{tab:comparison-modes}
\end{table*}

\subsection{Configurations} \label{game-config-struct}

\begin{figure}
    \centering
    \includegraphics[width=1\linewidth]{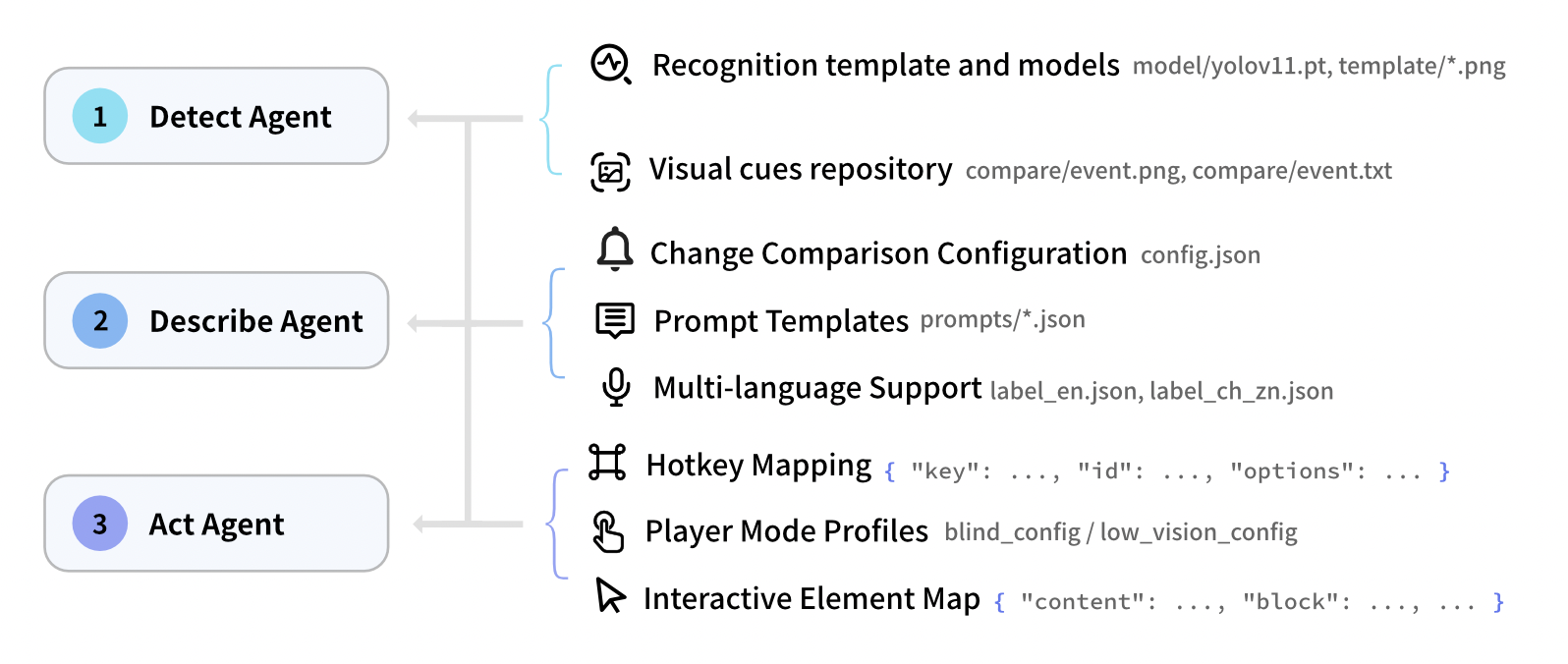}
    \caption{Structure of the configurations}
    \Description{This diagram details the structure of configurations for three agents: Detect Agent, Describe Agent, and Act Agent. The Detect Agent uses "Recognition template and models" (including files like model/yolov11.pt and template/.png) and a "Visual cues repository" (with files such as compare/event.png and compare/event.txt). The Describe Agent relies on "Change Comparison Configuration" (config.json), "Prompt Templates" (prompts/.json), and "Multi-language Support" (files like label_en.json and label_ch_zn.json). The Act Agent is configured with "Hotkey Mapping" (a JSON structure defining keys, IDs, and options), "Player Mode Profiles" (profiles for blind or low-vision users like blind_config), and an "Interactive Element Map" (a JSON structure with content and block details). These configurations enable each agent to perform their respective tasks, likely to support game interaction for users with visual impairments.}
    \label{fig:config-struct}
\end{figure}

The \textit{GamerAstra} framework employs a modular and scalable configuration system that supports the detect-describe-act working flow, with its configuration structure shown in Fig.~\ref{fig:config-struct}. This system enables game-specific accessibility functionalities to be deployed without requiring any modification to the game code or developer cooperation. All configurations are JSON-based and can be easily modified or extended by contributors without programming expertise.

\subsubsection{Detect Agent Configuration}

To enable the real-time event monitoring and visual analysis tasks of the \textbf{Detect} agent, the configuration system incorporates the following components:

\begin{itemize}[leftmargin=*]
    \item \textbf{Visual Cue Repository:} This directory stores cropped reference images and metadata for detecting critical in-game events. Each event is defined by a unique \texttt{event\_id} and represented by a pair of files (e.g., \texttt{event.txt}, \texttt{event.png}), where the text file contains description or reminder words. This minimal, self-contained structure enables contributors to add new event detectors by simply placing the corresponding files in a dedicated folder, without modifying any source code.

    \item \textbf{Recognition Models and Templates:} The \texttt{model/} folder contains pre-trained YOLOv11 model, while the \texttt{template/} folder stores PNG reference images for OpenCV-based template matching. The Detect agent executes a hybrid detection pipeline: template matching is used for static UI elements(e.g., UNO cards), while YOLO inference handles dynamic or rotating objects(e.g., mysuika fruits). Both pipelines share a common frame buffer, and detections are time-stamped to filter transient visual noise.
    
\end{itemize}

\subsubsection{Describe Agent Configuration}

The \textbf{Describe} agent generates context-aware auditory descriptions of visual elements and responds to player queries. Its behavior is governed by the following configuration files:

\begin{itemize}[leftmargin=*]

    \item \textbf{Change Comparison Configuration:}  
    The Change Comparison module's configuration files specify key parameters such as module enablement (e.g., \texttt{"enabled": true}), two similarity thresholds controlling sensitivity (e.g., \texttt{"threshold1": 0.3}, \texttt{"threshold2": 0.7}), and an optional list of UI regions to monitor (e.g., \texttt{"blocks": []}).  
    Such structured configuration enables the system to dynamically decide when to suppress insignificant changes, provide brief OCR-based feedback, or trigger detailed VLM-generated descriptions based on the visual similarity of consecutive frames.
    
    \item \textbf{Prompt Templates:} Located in \texttt{prompts/}, this module includes structured prompts for large language model (LLM) outputs, enabling the system to generate accurate, contextually grounded responses tailored to each game’s environment.
    
    \item \textbf{Multi-language Support:} All prompts, UI labels, and reminders support language-tagged variants(e.g., \texttt{label\_en}, \texttt{label\_zh\_cn}), allowing a single configuration set to support multilingual game interfaces without redundancy.
\end{itemize}

\subsubsection{Action Agent Configuration}

The \textbf{Action} agent facilitates non-visual execution of spatial interactions. Its functionality depends on the following configuration modules:

\begin{itemize}[leftmargin=*]
    \item \textbf{Interactive Element Maps:} The element maps are organized by game state. Each map entry is a merged result of:  (1) JSON outputs from OmniParser~\cite{lu2024omniparserpurevisionbased}, which processes full-screen game frames to produce structured UI element maps containing bounding boxes and semantic labels;  
    (2) manually defined interactive zones for elements not reliably detected automatically; and  
    (3) real-time OCR results to capture dynamic or text-heavy regions.  
    A typical entry follows the format \{\texttt{"block"}: [0.4273, 0.0985, 0.6030, 0.2125], \texttt{"content"}: "settings", \texttt{"interactivity"}: True\}, where \texttt{block} stores normalized coordinates $[x_1,y_1,x_2,y_2]$ for resolution-agnostic mapping, \texttt{content} holds the recognized label, and \texttt{interactivity} flags whether the element can receive input.
    
    \item \textbf{Hotkey Mappings:}  
    Hotkey configurations are stored under each game’s action module directory. Each hotkey entry defines key bindings (e.g., \texttt{"key": "<alt>+f"}), an action identifier (e.g., \texttt{"id": "get-current-score"}), and additional parameters such as target UI regions specified by normalized bounding boxes (e.g., \texttt{"options": {"block": [0.122, 0.053, 0.38, 0.21]}}) used by the agent to perform clicks.
    
    \item \textbf{Player Mode Profiles:} Distinct configuration files are provided for blind and low-vision users. These profiles define key interaction parameters, such as setting \texttt{input} variant to \texttt{keyboard} for blind users with grid-based navigation, or enabling \texttt{mouse} and \texttt{OCR} support for low-vision users. Runtime agents load these profiles to adapt feedback granularity, input methods, and interface behaviors accordingly.

\end{itemize}

\subsubsection{Generalizability and Easy Expansion}


By utilizing the configuration system, the GamerAstra framework is designed for easy expansion, with support for a wide range of games achieved through two complementary pathways: contributions from external configuration contributors and the \textit{Game Adapter} functionality. The \textit{Game Adapter} leverages LLMs to automatically generate game-specific adaptive prompts and temporary configurations for unsupported titles, enabling rapid integration and maintaining accessibility even for previously unseen games.


\textit{GamerAstra} supports several games by default, including:
\begin{itemize}[leftmargin=*]
\item \textit{Uno}: A strategic card game.
\item \textit{SenrenBanka}: A narrative-driven visual novel.
\item \textit{MySuika}: A casual 2D game.
\end{itemize}

In practice, configurations are expected to be enriched primarily through contributor efforts, while the \textit{Game Adapter} provides an automatic fallback to ensure baseline accessibility when such contributions are not available. This dual approach allows the framework to rapidly integrate new games while lowering the barrier for non-expert contributors to add or extend configuration files. We compare these modes in the technical evaluation~\ref{technical:comparison of modes}, and further details on the configuration ecosystem will be discussed as future work in the Discussion section~\ref{discussion:add more games}.

\subsection{Quantitative Performance Evaluation}

This section presents a quantitative evaluation of our framework, focusing on accuracy, coverage of user intent, and system performance across diverse gaming scenarios in three distinct gaming environments that we support: Uno, SenrenBanka, and MySuika. The analysis leverages empirical data from model testing, user interface navigation, and comparative assessments of different operational modes. 

\subsubsection{Detection Accuracy}

We quantitatively evaluated the accuracy of core detection agents within the GamerAstra framework via three games using three core components: Dialog Detection (YOLO), Object Recognition (YOLO in MySuika fruit detection), and Template Matching (Uno card detection). A total of 57 dialog screenshots that are not included in the original training set (45 from Google searches, 12 from the three supported games), 160 fruit images (covering 9 categories), and 119 Uno game-play frames were tested. Accuracy was measured by successful detection rates and false positives, and the result is as follow:

\begin{itemize}[leftmargin=*]
    \item \textbf{Dialog Detection}: Achieved 100\% accuracy on 12 game screenshots for supported games and 84.21\% overall accuracy across 57 images, demonstrating robust generalization to unseen games. False positives were eliminated through strict non-dialog labeling.
    \item \textbf{Fruits Recognition}: Achieved 97.6\% accuracy (160/164), while misclassifications (e.g. misclassify raspberry as an apple) occurred in 2 cases and another 2 wrong cases were caused by missing blueberries, which were very small.
    \item \textbf{Template Matching}: Attained 99.2\% accuracy (118/119) in Uno card detection, with a single failure caused by color and character similarity (Blue-8 misidentified as Blue-3).
\end{itemize}

The above validates our system's robustness for real-time gaming assistance, while strengths include low false positive rate and adaptability beyond static patterns. 

\subsubsection{Coverage of User Intent}

The supportability and executability of user intentions have always been key considerations. Therefore, we assessed intention coverage of our framework by testing navigation completeness and action execution in 45 different scenarios from: Uno (18), SenrenBanka (7), and MySuika (19). Tests evaluated whether all meaningful UI elements were detected and if user commands were correctly interpreted. The results are as follow:

\begin{itemize}[leftmargin=*]
    \item \textbf{UI Navigation}: 95.5\% success rate (42/44), with failures in Uno due to delayed OCR processing and residual template matches.
    \item \textbf{Action Execution}: 98.5\% success rate (66/67) in coordinate-based interactions. The failure case is due to inaccurate coordinates caused by fruit that is rotating too fast in MySuika. Other executions were all finished successfully.
\end{itemize}

GamerAstra demonstrates strong support for user intent, successfully identifying all UI navigation elements, responding seamlessly and executing user actions with high accuracy.

\subsubsection{Comparison of Modes}
\label{technical:comparison of modes}

To verify the importance of low-bar volunteer contributions to GamerAstra's configuration system, we classified the completeness of support for game configuration into three levels. Additionally, we included the NVDA~\cite{nvda} with OCR method as the baseline to conduct usability analysis across four levels in various gaming scenarios. The modes are shown below:

\begin{enumerate}[leftmargin=*]
    \item \textbf{NVDA + OCR}: Basic screen reader with OCR, acting as the baseline. As NVDA itself is not capable for direct interaction with the three games, OCR is the only existing method for game operation perception most of the time.
    \item \textbf{General Mode}: GamerAstra without any game-specific prompts and configurations.
    \item \textbf{Auto-adaptive General}: GamerAstra leverages the \textit{Game Adapter} for LLM-aided dynamic prompt generation and automated configuration enrichment, yet does not rely on human-aided annotations or configurations.
    \item \textbf{Full Framework}: GamerAstra with fully pre-annotated prompts and configurations specially customized for the games.
\end{enumerate}

To measure the four modes' ability for comprehension assistance and game operation assistance, we selected 15 scenes from the three games supported by default, including UI interfaces and playing scenes, to determine whether the four modes can help understand the game content and perform reasonable operations in these scenes. We define ``Understand'' as the cognitive capacity to assimilate all requisite informational elements necessary for comprehending a game scene's functional purpose, whereas ``Execute'' denotes the agent's ability to freely actualize all designer-intended interactions within said scene with full cognitive awareness. The result is shown in Table~\ref{tab:comparison-modes}.

Based on the results, the NVDA + OCR mode exhibits limited usability, while the General mode enhances usability in some interfaces, yet numerous non-operable elements persist due to the inability of VLMs to output accurate coordinates. The Auto-adaptive General mode is capable of generating more accurate descriptions and accelerates VLM descriptions through stored data; however, it does not fundamentally resolve the issue of non-operability, resulting in no increase in successful scenarios. The Full Framework, by utilizing manually annotated UI elements and game configurations, supports all scenarios, achieves high playability and avoids cognitive interference.

Moreover, our full framework can also reduce over-reliance on real-time VLM using saved cues, enhance the quality of the VLM's descriptions via game-specific prompts and filter irrelevant content (e.g. irrelevant animation effects in some parts of the game).

In summary, tailored configurations and pre-established contexts remain pivotal in supporting a broader spectrum of games at present. By leveraging the contributions of accessibility volunteers, the existing system is already capable of assisting BLV players in engaging with a variety of games.
\section{User Study}

To assess the effectiveness, usability and adaptability of our framework, we performed a user evaluation with 8 BLV individuals recruited through online BLV gaming chat groups (4 blind and 4 with low vision), ranging in age from 20 to 29 (mean = 23.25, SD = 3.2), and are referred to as $\text{U1}\sim \text{U8}$. U8 is female while $\text{U1}\sim \text{U7}$ are male players, as shown in Table.~\ref{tab:user-evaluation}. The study involved BLV participants performing structured gaming tasks across different game genres. The ``Visual Ability'' (Self-Reported Visual Ability) column indicates whether the player's visual impairment is congenital blind (BC), acquired blind (BA), low-vision that can recognize high-contrast content (LR) or low-vision that can detect light and shapes (LL).

\begin{table}[htbp]
\centering
\begin{tabular}{|c|c|c|c|c|}
\hline
\textbf{ID} & \textbf{Age} & \textbf{Gender} &\textbf{Visual Ability} & \textbf{Game Time} \\
\hline
U1 & 20 & Male & LL & 20h / week \\
U2 & 22 & Male & BC & 10h / week  \\
U3 & 29 & Male & LR & 5h / week \\
U4 & 21 & Male & LL & 8h / week \\
U5 & 25 & Male & BA & 15h / week \\
U6 & 20 & Male & BC & 20h / week \\
U7 & 26 & Male & BC & 10h / week \\
U8 & 23 & Female & LR & 5h / week \\
\hline
\end{tabular}
\caption{User Study Participants} \label{tab:user-evaluation}
\Description{The table titled "User Study Participants" presents information about 8 participants across five columns: ID, Age, Gender, Visual Ability, and Game Time. Each participant's data is listed as follows: U1 is a 20-year-old male with low-vision that can detect light and shapes, gaming 20 hours weekly; U2 is a 22-year-old male with congenital blindness, gaming 10 hours weekly; U3 is a 29-year-old male with low-vision that can recognize high-contrast content, gaming 5 hours weekly; U4 is a 21-year-old male with low-vision that can detect light and shapes, gaming 8 hours weekly; U5 is a 25-year-old male with acquired blindness, gaming 15 hours weekly; U6 is a 20-year-old male with congenital blindness, gaming 20 hours weekly; U7 is a 26-year-old male with congenital blindness, gaming 10 hours weekly; U8 is a 23-year-old female with low-vision that can recognize high-contrast content, gaming 5 hours weekly. The table summarizes demographic information including age and gender, visual ability categories without using abbreviations, and weekly game time for each participant in the user study.}
\end{table}

This study aimed to explore the following RQs:

\begin{itemize}[leftmargin=*]
    \item \textbf{(RQ1) System Usability}: How do BLV players perceive the usability of the GamerAstra framework?
    \item \textbf{(RQ2) Game Experience}: How does GamerAstra impact the gaming experience of BLV players?
    \item \textbf{(RQ3) Framework Granularity}: Does GamerAstra provide multi-granularity support for diverse games and player needs?
\end{itemize}

\subsection{Procedure and Metrics}

\subsubsection{Procedure}

Participants were randomly assigned to one of two experimental sequences using a Latin square design. NVDA~\cite{nvda}+OCR is established as the baseline, as it represents a method commonly employed by BLV players for experiencing games that lack accessibility support.

\begin{itemize} 

\item \textbf{Sequence A}: \begin{enumerate} \item Baseline phase using \textit{NVDA + OCR} with three games (\textit{Uno}, \textit{SenrenBanka}, \textit{MySuika}) \item Training on \textit{GamerAstra} framework (5--10 minutes) \item Experimental phase using \textit{GamerAstra} with the same games \end{enumerate}

\item \textbf{Sequence B}: \begin{enumerate} \item Training on \textit{GamerAstra} framework (5--10 minutes) \item Experimental phase using \textit{GamerAstra} with three games \item Baseline phase using \textit{NVDA + OCR} with the same games \end{enumerate}
\end{itemize}

The participants conducted the experiments by remotely connecting to our computers through RustDesk~\footnote{\url{https://rustdesk.com}}. All participants interacted with the same set of commercially available games lacking native accessibility support. The \textit{Qwen2.5-VL-Max} model~\cite{Qwen-VL} was deployed via Alibaba Cloud Model Studio API~\footnote{\url{https://www.alibabacloud.com/en/product/modelstudio}}. Each session lasted 1--1.5 hours, with participants receiving local currency equivalent to \$15 for compensation. Game presentation order was additionally randomized within phases to mitigate learning effects.

After completing the game-play sessions, participants were asked to complete a questionnaire evaluating their experience with both systems. In addition, the researchers conducted a short open-ended discussion to gather qualitative insights into participants' experiences, challenges, and overall impressions of GamerAstra.

\subsubsection{Analysis Metrics}

For system usability (RQ1), we assessed participants' perceptions of the system's ability to meet their requirements and its ease of use from the standardized UMUX-Lite~\cite{10.1145/2470654.2481287} questionnaire. For game experience (RQ2), the questionnaire examined whether GamerAstra improved game-play smoothness, reduced redundant or illogical information, and provided accurate spatial descriptions and task execution. For framework granularity (RQ3), we explored whether GamerAstra delivers differentiated, multi-granular support to meet the varied needs of BLV players across different game contexts. All items were rated on a 7-point Likert scale~\cite{Robinson2014}, with higher scores indicating more positive evaluations of the system's functionality.

\subsection{Analysis}

\begin{figure*}
    \centering
    \includegraphics[width=0.88\textwidth]{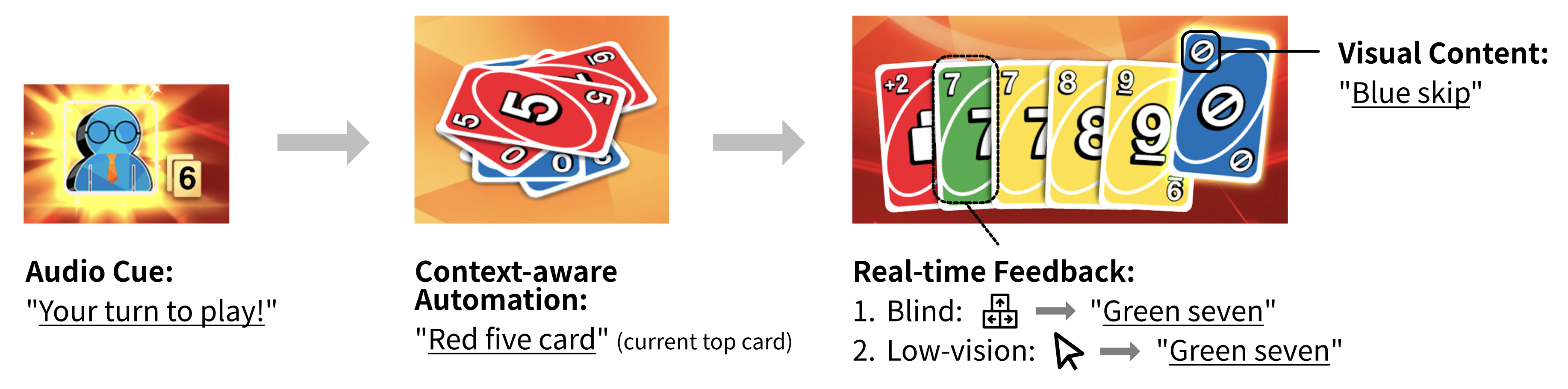}
    \caption{A Typical workflow for accessibility support in Uno, which represents RQ1.}
    \Description{This image illustrates a typical workflow for accessibility support in the Uno game, split into three sequential parts. First, an "Audio Cue" shows a player avatar with the text "Your turn to play!" to alert the user it's their turn. Next, "Context-aware Automation" displays Uno cards with the text "Red five card" (noting the current top card) to give context. Then, "Real-time Feedback" presents a hand of Uno cards: for blind users, a keyboard grid icon links to "Green seven," and for low-vision users, a mouse cursor icon also links to "Green seven." There's also "Visual Content" pointing to a blue skip card labeled "Blue skip." This workflow demonstrates how the system assists blind and low-vision users via audio cues, top-card context, and real-time card descriptions.}
    \label{fig:rq1-graph}
\end{figure*}

We analyzed interview transcripts and observational notes from eight BLV participants to gain deeper insights into how GamerAstra compared with the baseline. Thematic analysis revealed key differences in usability, game understanding, and personalized interaction across the two conditions.

\subsubsection{\textbf{RQ1: Usability of the system - Efficiency and accessibility perceptions}}

BLV participants consistently found that GamerAstra is significantly more usable than the baseline tool, particularly in reducing interaction complexity and improving accessibility. The baseline system often produces fragmented or non-sensical outputs (e.g., misidentifying curved text as garbled strings), forcing users to manually parse irrelevant content. U1, a low-vision user, criticized the native OCR for producing lengthy and illogical results, making games like \textit{Uno} unplayable, which was supported by U8. GamerAstra addressed these limitations through three core advancements, which is praised by U7, that our system ``effectively solves the information redundancy problem of OCR.'' A typical workflow is shown in Fig.~\ref{fig:rq1-graph}.

\textbf{Context-Aware Automation: } 
By integrating game logic, GamerAstra eliminated redundant steps. For example, in \textit{Uno}, the system directly announced played cards, eliminating the need for manual OCR scanning. U4 noted, \textit{``The interaction logic felt natural, no extra steps were needed.''}

\textbf{Real-Time Multi-Modal Feedback: } 
Participants praised features such as mouse-tracking vocalization, which dynamically read text under the cursor. U3 mentioned, \textit{``Real-time hover-based narration eliminated tedious screen scanning,''} and U5 appreciated that \textit{``Shortcuts delivered immediate, non-redundant feedback.''} Hybrid keyboard-mouse navigation was deemed critical, with U1 saying, \textit{``Horizontal navigation is indispensable in dense interfaces,''} while U4 described the interaction pattern as \textit{``intuitive and elegant.''}

\textbf{Adaptation to Visual Content: } 
The baseline OCR often struggled with interpreting non-text elements (e.g. images in \textit{MySuika}), rendering games \textit{``completely unusable''}. However, GamerAstra's logic-aware adaptations resolved this. U2, a blind participant, reported, \textit{``Positional cues allowed me to predict fruit collisions in MySuika, enabling strategic game-play.''}

\subsubsection{\textbf{RQ2: Game Experience – Enhanced Engagement and Cognitive Accessibility}}

\begin{figure}
    \centering
    \includegraphics[width=1\linewidth]{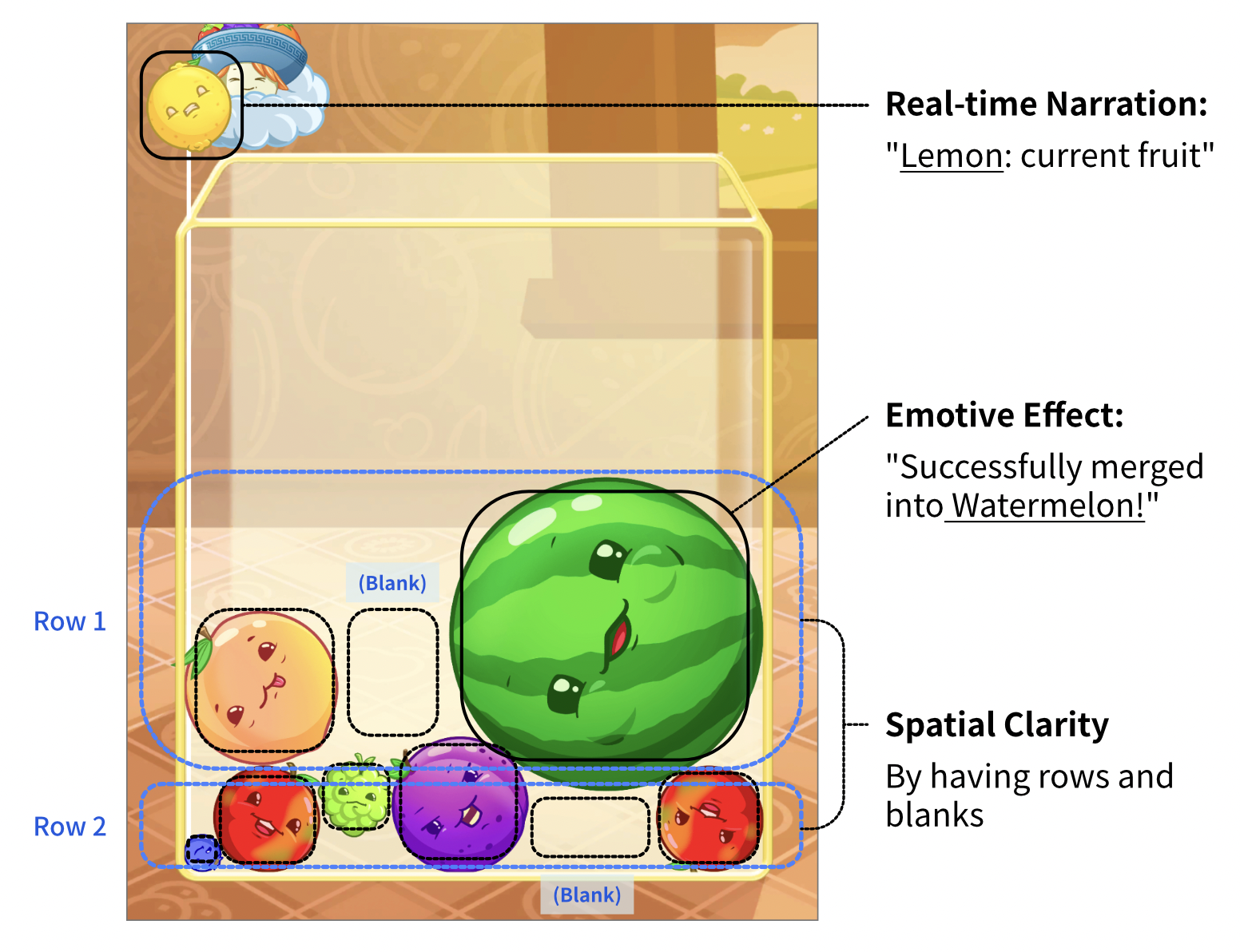}
    \caption{A Typical workflow for accessibility support in MySuika, which represents RQ2.}
    \Description{This image illustrates a typical workflow for accessibility support in the game MySuika, showing a game interface with anthropomorphic fruit characters arranged in rows (labeled Row 1 and Row 2, with blank spaces). Three accessibility features are highlighted: "Real-time Narration" points to a lemon character, stating "Lemon: current fruit" to describe the current game element; "Emotive Effect" points to a watermelon character, with the text "Successfully merged into Watermelon!" to convey a successful action; and "Spatial Clarity" explains how the arrangement of rows and blank spaces helps users understand the game’s layout. The scene demonstrates how the game provides real-time updates, emotional feedback, and spatial awareness to assist players, particularly those with visual impairments.}
    \label{fig:rq2-graph}
\end{figure}

GamerAstra significantly improved BLV players' ability to engage with games requiring spatial reasoning or narrative immersion, addressing two critical barriers.

\textbf{Spatial and Operational Clarity: }
In spatially complex games like MySuika, baseline OCR's inability to track dynamic elements left users \textit{``unable to infer object positions.''} GamerAstra's real-time spatial descriptions allowed U2 to \textit{``make strategy based on predicted collision outcomes''} by informing users of the fruit's position within the row and column, helping them establish a simple two-dimensional structure and understand the relative positioning. To further support spatial immersion, the system incorporated spatialized audio cues that conveyed the relative position of fruits on the screen, enabling users to perceive their approximate location and develop a clearer spatial understanding of the game environment.

\textbf{Narrative and Strategic Comprehension: }
For story-driven titles like SenrenBanka, baseline system ambiguities hindered engagement. GamerAstra's structured summaries clarified critical narrative elements. U6 remarked, \textit{``Dialogue options and character identities were disambiguated, eliminating guesswork.''} U2 added, \textit{``Narrative-driven games can provide real-time descriptions, which is more than enough for interaction. What we really need are real-time subtitles and scene descriptions, specifically for the interaction within the storyline.''}

\textbf{Emotional and Cognitive Impact: }
Participants reported reduced frustration and heightened confidence. U5 stated, \textit{``This framework restored my ability to play video games post-blindness, an empowering experience,''} while U6 linked usability to immersion: \textit{``Fluid interactions deepened my connection to the storyline.''} Meanwhile, U7 praised the "successful fruit merge" prompt in MySuika, noting it significantly boosted his sense of achievement.

\subsubsection{\textbf{RQ3: Framework Granularity – Adaptive Support for Diverse Needs}}

GamerAstra effectively provided multi-level assistance tailored to the diverse needs of BLV users across different game scenarios.

\textbf{Blind Users prioritized brevity}. U7 expressed, \textit{``I am very satisfied with the feedback. It meets my standards for concise descriptions and effectively addresses the issue of redundancy in OCR-only support.''} U5 concurred, noting that, "The framework's presentation of information aligns well with my preferences." Regarding additional descriptive prompts, our framework presents them using shortcuts favored by blind users to obtain extra information efficiently.

  
  
  
  
  
  

\textbf{Low-Vision Users leveraged on-demand queries} to supplement residual vision. \textit{``Active exploration via shortcuts compensated for my partial sight,''} said U3, who also remarked, \textit{``The real-time feature that tracks your mouse movement allows to directly announce the corresponding text, which is quite intelligent.''}
 
\textbf{Different games invoke distinct agent combinations.} U1 mentioned, \textit{``Although you are a framework, the intelligent adaptation you made for the three games is impressive.''} U2 also indicated, \textit{``I think this is a great idea. I can configure the profile for different game scenarios, and then the system will automatically play.''} For different game scenarios, we achieve targeted support through the framework's adaptability, allowing it to cater to the specific requirements of each game.

\subsubsection{Log Analysis}

\begin{figure*}[!hbt]
  \centering
  \includegraphics[width=\textwidth]{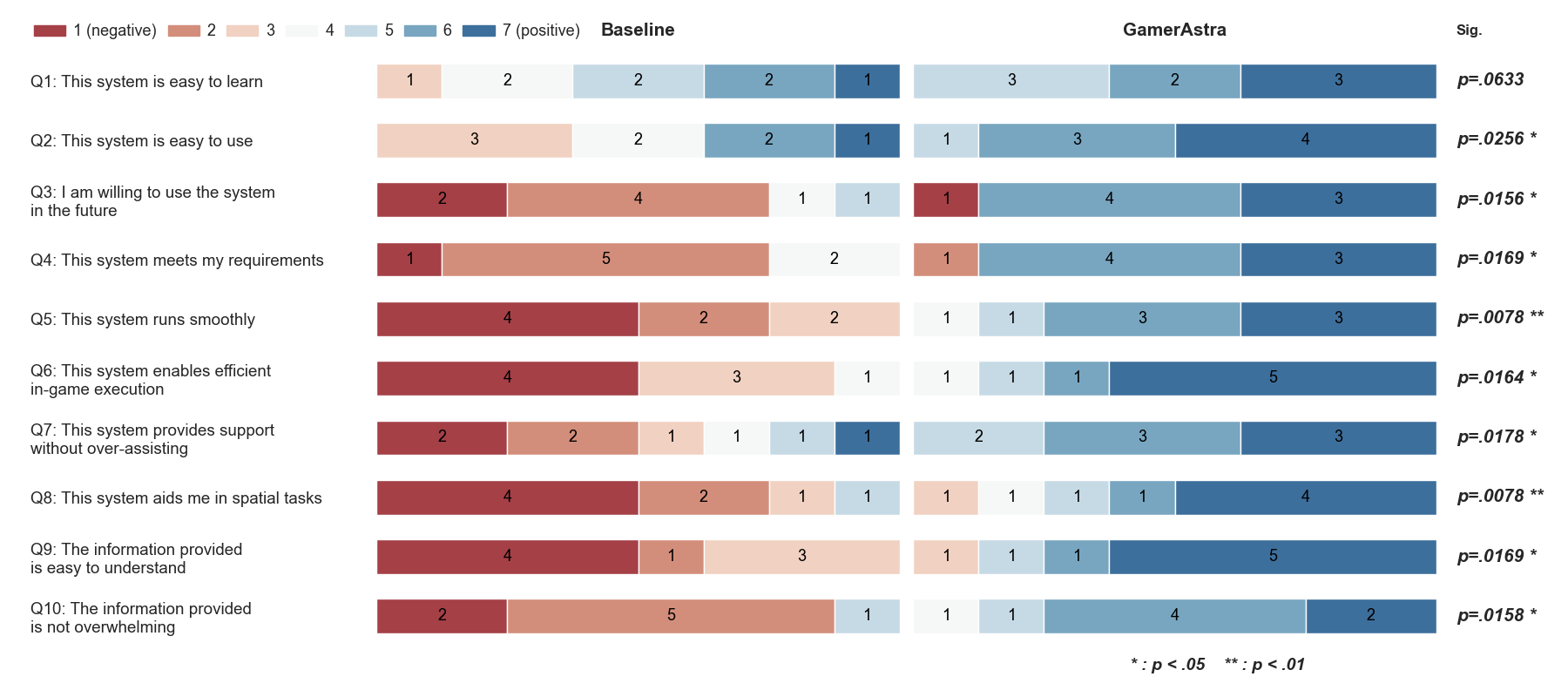}
  \caption{Distributions of the ratings for GamerAstra and the baseline (1=strongly negative, 7=strongly positive).}
  \label{fig:umux_distribution}
  \Description{This chart compares ratings for "GamerAstra" and a "Baseline" across 10 questions (e.g., ease of learning, use, willingness to use, system smoothness) using a 7-point scale (1=strongly negative, 7=strongly positive). The Baseline shows more negative (red/brown) ratings, while GamerAstra has more positive (blue) ratings. P-values indicate statistical significance: most questions (Q2, Q3, Q4, Q6, Q7, Q9, Q10) have one asterisk (p < .05), and Q5, Q8 have two asterisks (p < .01), meaning GamerAstra is rated significantly better in usability, performance, and information delivery compared to the baseline.}
\end{figure*}



We further analyzed system logs to objectively assess usability and efficiency. For each participant, we collected interaction records across gameplay sessions and categorized them by agent type. This analysis allows us to examine how GamerAstra provides multi-granular support (RQ3). 

Figure~\ref{fig:AGENTratios} presents the relative usage proportions of the three agents across different games, from which three key findings emerge:

\begin{figure*}[!hbt]
  \centering
  \begin{subfigure}[t]{0.48\textwidth}
    \centering
    \includegraphics[width=\textwidth]{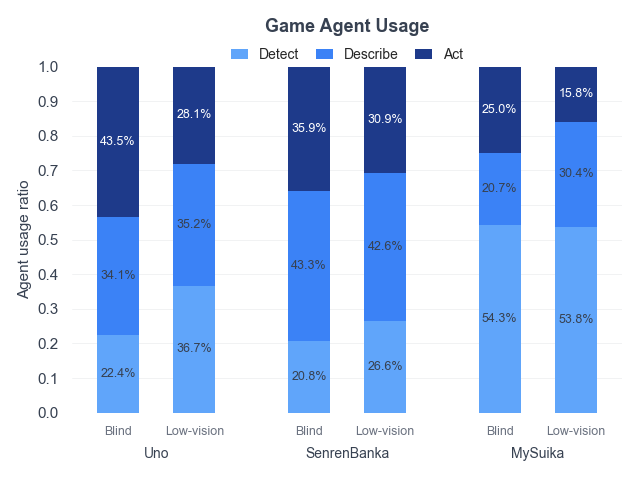}
    \caption{Agents usage ratios}
    \label{fig:AGENTratios}
  \end{subfigure}
  \hfill
  \begin{subfigure}[t]{0.48\textwidth}
    \centering
    \includegraphics[width=\textwidth]{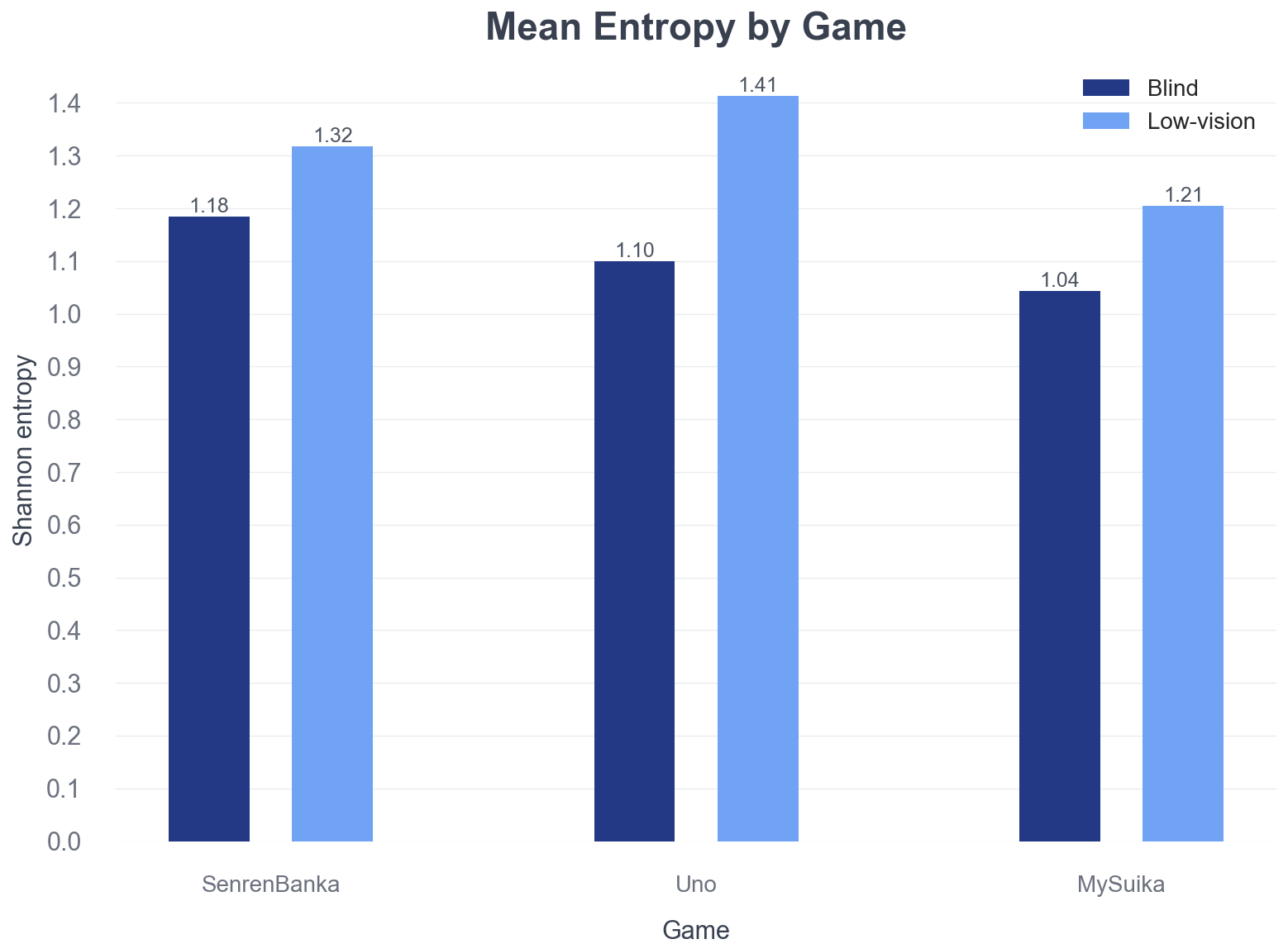}
    \caption{Entropy comparison between Blind and Low-vision groups}
    \label{fig:entropy}
  \end{subfigure}
  \caption{Comparison of agent usage ratios (left) and entropy measures (right) across different player groups}
  \Description{This image contains two bar charts comparing agent usage and entropy across player groups. The left chart (a) shows "Game Agent Usage" with stacked bars for "Act," "Describe," and "Detect" agents, split by Blind and Low-vision players in Uno, SenrenBanka, and MySuika (e.g., Uno's Blind group uses Act most, while Low-vision uses Detect most). The right chart (b) displays "Mean Entropy by Game," with Shannon entropy for Blind and Low-vision groups in each game—Low-vision players have higher entropy (e.g., Uno: 1.41 for Low-vision vs 1.10 for Blind). Together, they illustrate how agent usage and entropy differ between Blind and Low-vision groups across the three games.}
  \label{fig:combinedFigure}
\end{figure*}


\begin{itemize}
    \item \textbf{Blind users – Strong Act dominance}: Across all games, blind participants consistently exhibited a higher proportion of \textit{Act}-based operations, primarily through navigation and shortcut mechanisms. This demonstrates their reliance on direct, non-visual command execution to maintain gameplay efficiency. As U7, a blind participant, expressed: \textit{``This navigator helps me efficiently explore each element on the screen in a highly targeted way.''} 
    \item \textbf{SenrenBanka – Emphasis on Describe}: In the narrative-driven game \textit{SenrenBanka}, the \textit{Describe} agent emerged as the most active component, accounting for the largest share of interactions. This heightened descriptive activity reflects the game’s reliance on rich textual and contextual narration, requiring players to frequently invoke description functionalities in order to follow the evolving storyline. 
    \item \textbf{MySuika – Increased Detect activity}: In \textit{MySuika}, the \textit{Detect} agent was noticeably more active compared to other agents, showing the highest relative usage. The game’s fruit-placement mechanics involve rapid and frequent visual changes, which triggered sustained engagement with detection functionality to continuously recognize shifting scene states.  
\end{itemize}

To assess the diversity of player interactions, we employed the Shannon entropy~\cite{shannon} measure. For each participant group, we modeled the distribution of action frequencies during gameplay and computed the entropy as
\[
H(X) = - \sum_{i=1}^{n} p(x_i) \log p(x_i),
\]
where $p(x_i)$ denotes the probability of observing action $x_i$. Intuitively, higher entropy indicates a more even and diverse distribution of agent activities, while lower entropy reflects more concentrated or repetitive behavior.

Figure~\ref{fig:entropy} presents the mean entropy values across three games (\textit{Senrenbanka}, \textit{UNO}, and \textit{MySuika}) for both Blind and Low-vision player groups. Across all titles, the Low-vision group consistently exhibits higher entropy, suggesting a more balanced and diverse use of the three agents. This indicates that low-vision players employ a broader range of interaction strategies and display greater behavioral variability across agent types. As U1, a low-vision participant, noted: \textit{``The gameplay experience was great, there was no excessive assistance. I only used the framework when I needed help, which worked very well.''} In contrast, Blind participants show lower entropy values, reflecting more streamlined and command-driven patterns with repetitive reliance on particular agents. For example, they often repeatedly invoked the \textit{Describe} agent to compensate for overlapping auditory cues when background music coincided with in-game sound prompts.

Taken together, these findings highlight how levels of visual ability shape interaction preferences: low-vision players tend to seek support on demand, making more balanced use of multiple assistance channels, whereas blind players rely more consistently on descriptive feedback and keyboard-based controls. This underscores the importance of designing accessible interaction mechanisms that accommodate multiple levels of visual granularity, ensuring that both balanced and repetitive usage patterns are effectively supported.

To estimate a typical player's usage cost, we recorded the model call cost of each user in the experiment, which lasted $1\sim 1.5$ hour. On average, a player consumed 6425.8 input tokens and 331.6 output tokens, which costs \$0.0062 USD using Qwen2.5-VL-Max~\footnote{\url{https://www.alibabacloud.com/en/product/modelstudio}} (or \$0.019 USD using GPT-4o~\footnote{\url{https://openai.com/api/pricing/}}). This affordable price means a BLV player who plays video games 30 hours a week will consume only about \$0.45 (GPT-4o) every week. This is mainly due to the extensive use of pre-configured resources and local models like YOLO and OCR, which effectively reduce costs.

\subsubsection{Questionnaire Analysis}

In the questionnaire study of the 8 users, based on the three identified RQs, we included two all-positive UMUX-Lite items (Q2, Q4) along with several self-defined questions, all rated on a 7-point Likert scale. Given the well-established correlation between UMUX-Lite and system usability scale (SUS) scores~\cite{umux_article, 10.1007/978-3-319-20886-2_20, 10.5555/2993219.2993221}, we computed SUS-equivalent scores for both systems using the conversion formula proposed in~\cite{10.1145/2470654.2481287}, as shown in Equation~\ref{eq:sus-conversion}. The Questionnaire results are summarized in Fig.~\ref{fig:umux_distribution}.
\begin{equation}
\text{SUS Score} = 0.65 \times \left( \frac{\sum_{i=1}^{2} \text{UMUX-Lite}_{i} - 2}{0.12} \right) + 22.9 \pm 1.1
\label{eq:sus-conversion}
\end{equation}

The SUS result reveals clear improvements in GamerAstra compared to the baseline system. GamerAstra achieved a SUS score of $78.42\pm1.1$, which significantly outperformed baseline ($49.31\pm1.1$), demonstrating enhanced usability. Subsequently, we employed the Wilcoxon Signed-Rank Test~\cite{Rey2011} to evaluate the improvements. The resulting test statistics and $p$-values are presented in Fig.~\ref{fig:umux_distribution}.

Based on user ratings and Wilcoxon signed-rank tests, GamerAstra demonstrates statistically significant improvements over the baseline across multiple dimensions of usability (RQ1) and immersion (RQ2). The rightward shift in rating distributions Fig.~\ref{fig:umux_distribution} from predominantly neutral / negative (light red) in baseline to positive (light / dark blue) in GamerAstra—visually confirms these advancements. Notably, spatial task support ($p = 0.0078$), smooth experience ($p = 0.0078$), usage willingness ($p = 0.156$), information clarity ($p = 0.158$), efficient execution ($p = 0.164$) showed the most pronounced upgrades, aligning with the SUS score disparity. Overall, the results underline the system's significant success in combining functional performance with user-centered design.
    
\section{Discussion} \label{discuss}

Our paper identifies key challenges for the BLV community in video gaming support and demonstrates the potential of a multi-agent approach to enhance the accessibility of 2D video games. We will outline the next steps below.

\subsection{Exploring the Enhancement of Accessibility Features}

\subsubsection{Improving existent accessible games}
Currently, our framework is only tested and utilized for games completely without accessibility support. However, current accessibility support is also not satisfactory in several scenarios~\cite{ran2025usersblindlowvision, martinez2024playing}, limiting their gaming experience. Therefore, it will also be beneficial for them to also use the GamerAstra framework as a supplement. Moreover, some existing accessibility modules also fail to satisfy low-vision players~\cite{othman2023serious, ran2025usersblindlowvision}, forcing users to either endure inadequate adaptations or abandon assistive tools altogether, which underscores the necessity for customizable, multi-tiered visual assistance systems, and GamerAstra may also resolve this problem. Future work may focus on integrating GamerAstra with existing accessibility modules to enhance their functionality, ensuring seamless compatibility and user-centric design, which makes U5 and U6 extremely excited about it. U6 reported that as a blind person, he was unable to fully interact with certain elements in \textit{Ace Attorney} and \textit{The Invisible Guardian}, preventing him from completing these games. He hopes to use our tool to better explore them in the future.

\subsubsection{Emotive design}
BLV individuals rely on generic, emotionless text-to-speech services to interact with computers and our framework, even during game sequences where emotions should be richly expressed, although the sound effects in the game can still enhance the atmosphere to some extent. Therefore, emotive design that includes emotionally modulated synthetic voices and affective narration~\cite{yang2025emovoice, cho2024} may further improve the user experience. This approach can serve as an alternative to the emotional engagement and aesthetic ambiance that sighted players typically derive from visual elements to improve the emotional experience for BLV players.

\subsubsection{Enhancing software}
In addition, although screen readers can assist visually impaired users in accessing many software applications, there are still numerous programs with inaccessible sections and interfaces, as reported by user study participants. Among these, icon-only buttons further hinder accessibility via OCR. The current solution often involves developing specialized screen reader plugins for optimizing these software applications, such as NVDA's \textit{WhatsAppPlus}, \textit{WeChatEnhance}, and \textit{eMule} plugins. However, the relatively high technical barrier to plugin development still limits better support for more software. Future research should also investigate GamerAstra's solutions for these scenarios, as the underlying logic for both gaming support and software support is likely to be interconnected.

\subsection{Potential for Supporting More Complex Gameplay and Twitch-based Interactions}
\label{discussion:add more games}

The current framework demonstrates effectiveness in small and medium-sized games, yet many modern games with open-world environments, intricate gaming mechanics and hundreds of roles and items are more complex. Since configuring these games requires more work, we have not conducted exhaustive tests yet. To address this, a \textit{crowd-sourced ecosystem} could be further developed, where contributors collaboratively annotate game content (e.g. mapping terrain features, documenting roles and items' properties, or scripting combat strategies) through an online platform, as exemplified by \textit{YouDescribe}~\cite{Pitcher-Cooper2023May}, a crowd-sourced video description platform that helps BLV people. The platform can also further accelerate prompt engineering using AI techniques. Such a no-code crowd-sourcing model may contribute to the democratization of accessibility, facilitate the customized development of accessibility features and accelerate the support for indie games and games from the Global South. However, future research should also further investigate another underexplored issue: whether similar accessibility crowd-sourcing platforms might discourage developers from improving their own platforms' accessibility performance.

Meanwhile, \textit{reinforcement learning methods} may also emerge as a promising future research direction. For instance, AppAgent and AppAgentX proposed a reinforcement learning-based approach for autonomously exploring UI elements and constructing knowledge bases~\cite{zhang2025appagent, jiang2025appagentx}. Although this method was originally proposed to support autonomous GUI agent, this methodology shows potential for further applications in automated recognition and annotation of game content, possibly serving as an ``intelligent tour guide'' for BLV players during game-play. Should gaming agents~\cite{tan2024cradle, chen2024vlmsplayactionroleplaying} advance further, subsequent research could investigate the use of in-game agents to assist players' actions in harder sections, thereby emulating the presence of a sighted co-player.

Moreover, \textit{path-finding in complex maps} and \textit{decision support in twitch-based games}, which are not yet supported by the current framework, remain challenging using current technologies and require future work due to the limitations of VLM backbones and traditional computer vision techniques in spatial perception and complex spatial understanding, as well as the high reasoning latency of VLM-based methods~\cite{xu2024gameagents}. This might become an important direction for future research. For example, U2 showed a strong desire for this feature during the evaluation. 

\subsection{Towards a Unified Gamer Society}

During the user study, participants expressed enthusiasm about the potential of GamerAstra to bridge the gap between BLV players and the broader gaming community. U6 reported satisfaction in being able to experience well-known titles such as \textit{SenrenBanka} by Yuzusoft~\footnote{\url{https://www.yuzu-soft.com}}, a prominent game producer. He emphasized that gaining access to such popular games would allow him to share common topics of discussion with sighted friends, reducing social isolation and fostering inclusivity, as also supported by prior research~\cite{gonccalves2020playing}. Similarly, U5 appreciated the framework’s support for \textit{Uno}, noting that BLV players often remain restricted to a limited selection of specially designed board games with small player bases. He indicated that although there are accessible Mahjong games and a local poker card game with good accessibility support, these versions are not the ones popular among sighted players, which lack accessibility features. Therefore, he expressed a strong expectation for effort in making such games accessible, which would enable BLV users to engage with and join competition with global player communities numbering in the millions.

These responses highlight a broader social need: accessible gaming is not solely about individual playability, but also about participation in shared cultural experiences. When BLV players can access the same games as their sighted peers, it helps dissolve barriers and promotes a more unified gaming society. Future work should therefore prioritize expanding support to high-profile and multiplayer games, ensuring that accessibility does not come at the cost of exclusion from popular digital social spaces.
\section{Conclusion}


We introduced \textbf{GamerAstra}, a multi-agent AI framework designed to enhance the accessibility of 2D non-twitch video games for BLV players, eliminating the need for programming skills or official developer support. GamerAstra addresses these challenges by employing a multi-agent system that can help people to adapt new games to be accessible, provides tailored support for various levels of visual impairment, enhances navigation support, and integrates agent feedback to assist in mental model construction. Our evaluation, which includes comparative user studies and quantitative assessments, demonstrates that GamerAstra significantly improves gaming accessibility and user experience. These results underscore the potential of GamerAstra to transform gaming experiences for BLV players, making games more inclusive and enjoyable.

\section*{Acknowledgments}
We acknowledge the partial use of a large language model, specifically ChatGPT, to assist in the writing process. The LLM was employed as a tool for polishing the manuscript to enhance the clarity and quality of the text.

\bibliographystyle{ACM-Reference-Format}
\bibliography{CHI/reference}


\end{document}